\documentclass[useAMS,usenatbib]{mn2e}
\usepackage{color}
\usepackage{graphicx}
\usepackage{array}
\usepackage{color}
\usepackage{comment}
\usepackage[utf8]{inputenc}  %Additional package for equation
\usepackage{amsmath,breqn}  %addition package for math
\usepackage{natbib}
\bibpunct{(}{)}{;}{a}{,}{,}

\newcommand \mypath{{./figures}}
\newcommand{\degree}{\ensuremath{^\circ}}

\usepackage[official]{eurosym}
\usepackage{amssymb}
\usepackage{float}
\usepackage{tabularx}
\usepackage{footnote}
\usepackage{booktabs}
\usepackage[T1]{fontenc}
\title[Differential interferometry of QSO BLRs]{Differential interferometry of QSO broad line regions I: \\
       improving the reverberation mapping model fits and black hole mass estimates}
\author[S. Rakshit et al.]{Suvendu Rakshit$^{1}$\thanks{email:{suvendu.rakshit@oca.eu and suvenduat@gmail.com}},
 Romain G. Petrov$^{1}$,
 Anthony Meilland$^{1}$,
 Sebastian F. H\"onig$^{2}$ \\
 $^{1}$Laboratoire Lagrange, UMR 7293, University of Nice Sophia-Antipolis, CNRS, Observatoire de la Cote D’Azur, \\BP 4229, 06304, Nice Cedex 4, France.\\
$^{2}$Dark Cosmology Center, University of Copenhagen, Juliane Maries Vej 30, 2100 Copenhagen, Denmark\\}

\begin{document}
\date{Accepted------}

\pagerange{\pageref{firstpage}--\pageref{lastpage}} \pubyear{--}

\maketitle

\label{firstpage}

\begin{abstract}
Reverberation mapping estimates the size and kinematics of broad line regions (BLR) in Quasars and type I AGNs. It yields size-luminosity relation, to make QSOs standard cosmological candles, and mass-luminosity relation to study the evolution of black holes and galaxies. The accuracy of these relations is limited by the unknown geometry of the BLR clouds distribution and velocities. We analyze the independent BLR structure constraints given by super-resolving differential interferometry. We developed a three-dimensional BLR model to compute all differential interferometry and reverberation mapping signals. We extrapolate realistic noises from our successful observations of the QSO 3C273 with AMBER on the VLTI. These signals and noises quantify the differential interferometry capacity to discriminate and measure BLR parameters including angular size, thickness, spatial distribution of clouds, local-to-global and radial-to-rotation velocity ratios, and finally central black hole mass and BLR distance. A Markov Chain Monte Carlo model-fit, of data simulated for various VLTI instruments, gives mass accuracies between 0.06 and 0.13 dex, to be compared to 0.44 dex for reverberation mapping mass-luminosity fits. We evaluate the number of QSOs accessible to measures with current (AMBER), upcoming (GRAVITY) and possible (OASIS with new generation fringe trackers) VLTI instruments. With available technology, the VLTI could resolve more than 60 BLRs, with a luminosity range larger than four decades, sufficient for a good calibration of RM mass-luminosity laws, from an analysis of the variation of BLR parameters with luminosity.

\end{abstract}


\section{Introduction}

Active galactic nuclei (AGNs) are powered by the accretion of matter onto a supermassive black hole (SMBH). The high energy photons from the accretion disk illuminate gas clouds and dust clumps. According to the standard model of AGNs \citep{1993ARA&A..31..473A,1995PASP..107..803U}, the broad emission lines are produced by an inner region of high velocity gas clouds called the broad line region (BLR). In Seyfert 2 AGNs, an edge-on clumpy dust torus shields the BLR. To understand QSOs we must characterize the geometrical and velocity distribution of the BLR clouds. This yields the mass and accretion rate of the central SMBH, and the mechanisms to transport material inwards to the accretion disk and outwards to the jet. It constrains the distribution of light emitted by the accretion disk, and hence constrains the accretion disk models.

So far, the only technique providing some direct constraints on the BLR geometry has been the so-called Reverberation Mapping \citep[RM;][]{1982ApJ...255..419B,1993PASP..105..247P}. Early RM observations have led to estimate the BLR characteristic size, providing a size-luminosity relation, but have failed to directly constrain their actual geometry \citep{2000ApJ...533..631K,2013ApJ...767..149B}. They have nevertheless provided mass estimation for the SMBH themselves used to establish an empirical mass-luminosity relation for SMBHs \citep{2004ApJ...613..682P,2006ApJ...641..689V,2009ApJ...705..199B,2010ApJ...708.1507B}. With improved precision, the size-luminosity relation would be an useful tool for cosmology, making QSOs usable as standard candles to independently estimate distances, and the mass-luminosity relation will constrain the global evolution of QSOs and host Galaxies  \citep{2011ApJ...740L..49W, 2013ApJ...767..149B}.

Spatially resolved observations of QSOs have been a major goal of optical and infrared long baseline interferometry (OI) since its origin. Interferometric observations of QSOs have however proven difficult in practice: their apparent magnitude makes them particularly hard to observe with interferometry. In addition, the BLR characteristic sizes deduced from RM observations suggest typical angular diameters smaller than 0.3 mas, i.e. about ten times below the standard resolution limit of the largest infrared interferometers. \citet{2001CRPhy...2...67P} and \citet{2003Ap&SS.286..245M} have proposed to apply to BLRs the
super-resolution capability of differential interferometry \citep {1989dli..conf..249P}. This was one of the design specifications of the VLTI first generation instrument AMBER \citep{2007A&A...464....1P}. Because of the unexpectedly low performances of fringe stabilization at the VLTI, \citet{2012SPIE.8445E..0WP} introduced a new observation and data reduction technique that boosted the limiting magnitude of VLTI/AMBER in medium spectral resolution and allowed the first successful differential interferometry observation of the BLR of 3C273. This success and the need to explain the data triggered the development of the geometric and kinematic model described in this paper. We use it to evaluate the contributions of spectro-interferometry with the VLTI, combined with RM, to discriminate between different BLR geometries, to constrain the key geometrical and kinematics parameters and finally to estimate the BH mass.

In section 2, we overview the two BLR observing techniques reminding how their key observables are computed. In section 3, we describe our BLR model and show how it allows estimating RM and OI observables. In section 4, we display and discuss the RM and OI signatures of some key model characteristics. In section 5 we compute the number of targets accessible to the VLTI with various existing, near future or possible instruments. We estimated the accuracy on the parameters that can be achieved with current OI in section 6, and section 7 gives a final synthetic discussion of the potential of OI with the VLTI to confront, check and extend the possibilities of RM, improve the mass luminosity relation and the use of QSOs as standard mass tags.

\section{Observing techniques}
%Here we summarize the principle, achievements and the limitations of RM as well as the OI interferometric measures and its applicability to the unresolved sources like BLR.
 \subsection{Reverberation Mapping}\label{sec:RM}
 
 Reverberation mapping (RM) is based on the spectro-photometric variability study of the continuum and line flux \citep{1982ApJ...255..419B,1993PASP..105..247P}. A part of the continuum light emitted from the central compact source travels directly toward the observer. A fraction of this light is absorbed by the BLR clouds and emitted in a narrow spectral bin depending on the specific Doppler shift of each cloud. This echo of light is detected at a wavelength characterizing the radial velocity of the cloud with a time delay characterizing its distance to the central source as well as its relative distance to the observer. If $C(t)$ is the light curve of the continuum and $L(v, t)$ is the flux detected in the line at the radial velocity $v$ and time $t$ we can write:
  \begin{equation}  
  L(v, t) = \int\limits_{-\infty}^{\infty}{\Psi(v, \tau)C(t-\tau)d\tau},  
  \label{eq:LVT}                                      
  \end{equation}
  where $\Psi(v, \tau)$ is the 2D probability density function of velocities and time lags for all BLR clouds. It is often called the RM echo function or the RM  transfer function \citep{1982ApJ...255..419B,1991ApJ...379..586W}. A very accurate and well sampled $\Psi(v, \tau)$, de-convolved from local line shapes and combined with physical constraints can in principle constrain almost uniquely the BLR geometry and velocity field. 
  
  However, because of sparse, undersampled and noisy data, most of early RM work has been based on the measurement of the typical time lag $\tau_{\mathrm{cent}}$ and dynamical line width $\Delta V$. The time lag $\tau_{\mathrm{cent}}$ is the barycenter of the 1D transfer function $\Psi(\tau)$ resulting from the integration of $\Psi(v, \tau)$ over $v$. For infinite time series, $\tau_{\mathrm{cent}}$ is equal to the centroid of the cross-correlation $CCF(\tau)$ between $C(t)$ and $L(t)$ where $L(t)$ is the integration of $L(v, t)$ over $v$ \citep{2001sac..conf....3P}. The approximate width $\Delta V$ of the 1D dynamical line profile $\Psi(v)$ ($\Psi(v, \tau)$ integrated over $t$), is given by FWHM or the standard dispersion $\sigma_l$ of the “mean” or the “rms” emission line profile \citep{2006A&A...456...75C}.

 A simple virial relation \citep{2004ApJ...613..682P} links the $\Delta V$ and $\tau_{\mathrm{cent}}$ to the mass of the central SMBH: 
  \begin{equation}  
  M_{\mathrm{bh}}=f \frac{R_{\mathrm{blr}}{\Delta V}^2}{G}, 
  \label{eq: Mass}                                         
  \end{equation}
  where $R_{\mathrm{blr}}=c\tau_{\mathrm{cent}}$ is the RM BLR size, $G$ is the gravitation constant, $c$ is the speed of light and $f$ is an unknown scale factor that depends mainly on the BLR geometry and kinematics. This relation allowed to estimate SMBH masses that can be related to the luminosity of the BLR:  $M_{\mathrm{bh}} \propto L^{0.79 \pm 0.09}$ \citep{2004ApJ...613..682P, 2006ApJ...641..689V,2009ApJ...705..199B}. However the statistical scatter of masses in this relation is larger than a factor 3 for all the estimated masses by \citet{2000ApJ...533..631K} \citep{2006A&A...456...75C}. 
  
  %For specify targets, such as 3C273, the mass scatter is even larger, ranging from $0.235^{+0.037}_{{}-0.033}  \times 10^9M_\odot$ in \citet{2000ApJ...533..631K} to $6.59^{+1.68}_{{}-0.9} \times 10^9 M_\odot$ in \citet{2005A&A...435..811P}.
  
  Many efforts have been made to improve this mass-luminosity relationship that is vital to study the evolution of mass and accretion rate of SMBHs as a function of their age (via the redshift), and the co-evolution between AGNs and host galaxies. Some authors improved the measurement of the time lag $\tau_{\mathrm{cent}}$ interpolating the light curves with a damped random walk model \citep{2009ApJ...698..895K,2011ApJ...735...80Z}, if possible simultaneously through several lines with different scale factors. Others have tried to improve the estimation of the projection factor $f$ from model fits of the line profile and the time cross-correlation function \citep{2001ApJ...551...72K,2004ApJ...615..645O,2006A&A...456...75C,2012MNRAS.426.3086G}. They show that in relatively flat models, such as disks \citep{2001ApJ...551...72K} or bowl geometry \citep{2012MNRAS.426.3086G} the main parameter affecting $f$ is the global inclination $i$ of the BLR. The estimation of $\Delta V$ from the “mean” and “rms” spectrum depends on the relative contributions to the line width of the global velocity field, such as global rotation, inflow or outflow, and more local contributions, such as microturbulence inside the clouds as well as the macroturbulent motion of the clouds themselves. 
%Some recent progress have been made by models parameterizing macroturbulence \citep{2006A&A...456...75C,2012MNRAS.426.3086G,2014MNRAS.445.3055P}

 %Estimations of the  from “mean” and “rms” spectrum indirectly deal with that problem that has also been tackled by direct model fitting of models parameterizing macroturbulence \citep{2006A&A...456...75C,2012MNRAS.426.3086G}.
 
 Another important achievement of RM is the size-luminosity, i.e. lag-luminosity relationship. In spite of all efforts to improve the time lag estimation \citep{2009ApJ...698..895K,2011ApJ...735...80Z} as well as to isolate the BLR luminosity from the host galaxy \citep{2006ApJ...644..133B}, the dispersion of lags around the best fit of the lag-luminosity relation is between 0.13 and 0.21 dex \citep{2013ApJ...767..149B} while an accuracy better than 0.05 dex would be necessary to allow the abundant QSOs to be as good standard candles as the scarce Type Ia supernova. 
  
Recent RM high quality datasets allowed some successful reconstructions of $\Psi(\tau, v)$ and the detection of inflow and outflow signatures \citep{2010ApJ...720L..46B,2013ApJ...764...47G}. Direct modeling of RM data using a Bayesian inverse problem approach improves the number and accuracy of fitted parameters \citep{2011ApJ...730..139P,2011ApJ...733L..33B,2013arXiv1311.6475P,2014arXiv1407.2941P}. Additional observational constraints are however necessary, to independently remove degeneracies of model parameters and improve the accuracy of their fit. We will show that a good candidate is optical interferometry that can provide useful angular and dynamic constraints on unresolved sources when it is used in its spectro-interferometry or differential interferometry mode.

 %###########################################  Optical Interferometry #######################################
 %###########################################  ====================  ###########################################
 
 \subsection{Optical interferometry}\label{sec:interferometery}

Multi-telescopes optical interferometry (OI) is intended to provide very high angular resolution information, and ideally images, with spatial (or angular) resolution $\sim \lambda/B$ where $\lambda$ is the observation wavelength and $B$ is the interferometer baseline, i.e. the maximum distance between apertures.
Resolved observations of AGNs have been a goal for OI since its redefinition by Labeyrie in the 70s \citep{1978ARA&A..16...77L,1986A&A...162..359L} but these targets are generally too faint for most existing facilities, except for low spectral resolution observations in the near- and mid-infrared, with 8-10 meter aperture telescopes, only available at the VLTI and the Keck interferometer (KI).
% The resolution of AGNs has been a goal for OI since its redefinition by Labeyrie in the 70s \citep{1978ARA&A..16...77L,1986A&A...162..359L} but these targets have been too faint for all optical interferometers, except for near and mid infrared low spectral resolution observations with the VLTI with 8-m UTs and the KI with 10-m telescopes.
Since \citet{2004Natur.429...47J} more than 45 AGNs have been successfully observed in the $N$ and $K$ bands. This allowed to constrain the size of the innermost dust torus structure and revealed its complexity. \citet{2013A&A...558A.149B} reject the existence of a simple size-luminosity relation in AGNs, because the $L^{0.5}$ scaling of bright sources fails to properly represent  fainter sources. There are clearly several components, with at least a cooler more equatorial structure and a hotter more polar one. However \citet{Kishimoto2014} still tries to find an unification scheme based on the idea that in low luminosity AGNs the inner torus is more shallow than in high luminosity ones, because a latitudinal dependent radiation pressure blows away all material far from the equatorial plane in high luminosity AGNs. Thus low luminosity AGNs would have much more dust clouds in the polar direction. Both the KI and the VLTI measurements, summarized in \citet{2012JPhCS.372a2033K}, show that in the $K$-band, the dust torus inner rim size is fairly close to a $R_{\mathrm{rim}} \propto L^{0.5}$ size as first indicated by the infrared RM observations of \citet{2006ApJ...639...46S}, with a size excess with regard to $\propto L^{0.5}$ that increases as $L$ decreases but remains small in the $K$-band. In section §\ref{sec:present and future} we will use this Suganuma size as a lower limit of the inner rim size to estimate the feasibility of AGN OI observations.
  
 The OI observation of AGN BLRs faces two major difficulties. First, a spectral resolution greater than 500 is required, which impacts the sensitivity of the
 instrument and places AGNs out of reach of the first generation VLTI instruments. A solution would be to assist the instrument with a fringe tracker (FT) that would enable the use of exposure time greater than the atmospheric coherence time for differential piston. As the limiting magnitude of current fringe tracker remains lower than $K=9.5$, we developed a new observing and data processing technique, based on the visible OI data processing by \citet{1999JOSAA..16..872B}, that allows to observe fainter sources in medium and high spectral resolution without fringe tracker. This allowed the first successful observation of the BLR of 3C273 \citep{2012SPIE.8445E..0WP}.
 Second, the sizes deduced from RM observations suggest typical angular diameters for BLRs to be much smaller than the formal angular resolution of current facilities observing in the $K$-band, and we have to use the super resolution power of differential spectro-interferometric measurements.
 %First, we need to observe at a spectral resolution typically larger than 500 and this lowers the sensitivity limit of the instrument and places all AGNs out of reach, except if the science instrument is assisted by a fringe tracker (FT) allowing individual exposure time longer than the atmosphere coherence time for differential piston. As the limiting magnitude of current fringe tracker remains lower than $K=9.5$, we developed a new observing and data processing technique, based on the visible OI data processing by \citet{1999JOSAA..16..872B}, that allows to observe fainter sources in medium and high spectral resolution without fringe tracker. This allowed the first successful observation of the BLR of 3C273 \citep{2012SPIE.8445E..0WP}. Second, from RM sizes, we expect all BLRs to be much smaller than the angular resolution of the VLTI and KI in the $K$-band. However, the spectro-interferometric instruments like AMBER allow accurate differential interferometry measures on non-resolved sources.
  
 %\bibitem[Berio et al.(1999)]{1999JOSAA..16..872B} Berio, P., Mourard, D., Bonneau, D., et al.\ 1999, Journal of the Optical Society of America A, 16, 872 
 
 \subsubsection{Spectro-interferometric measurements}\label{sec:specto-interferometery}
 
  \citet{2007A&A...464....1P} provide a thorough description of the spectro-interferometric instrument AMBER and its uses. This section offers a concise summary of the concepts of interferometry that are of direct relevance to the application presented in this paper. An interferometer with baseline $B$ yields the complex visibility of the source, i.e. the normalized Fourier Transform $\tilde{O}(\mathbfit{u}, \lambda)$ of the source brightness distribution $O(\mathbfit{r}, \lambda)$ at the spatial frequency $\mathbfit{u}=\mathbfit{B}/\lambda$
  %{\small
  \begin{multline}
  	\tilde{O}(\mathbfit{u}, \lambda) =\frac{\int\int{O(\mathbfit{r}, \lambda)} \mathrm{e}^{-2\pi \mathrm{i} \mathbfit{u.r}}\,\mathrm{d^2}\mathbfit{r}}{\int\int{O(\mathbfit{r}, \lambda)\, \mathrm{d^2}\mathbfit{r}}} = V_{*}(\lambda) \mathrm{e}^{\mathrm{i}\phi_{*}(\lambda)}.	
  \label{eq:OI}
  \end{multline} %}
  The modulus $V_{*}(\lambda)$ of $\tilde{O}(\mathbfit{u}, \lambda)$ is given by the contrast of the fringes and called source absolute visibility. It needs to be calibrated on a known reference source, which limits its accuracy. The position of the fringes yields the phase $\phi_{*}(\lambda)$ of the source complex visibility. It can be measured only with regard to some internal phase reference or iteratively constrained by closure phase relationships of triplets of baselines when at least three
  baselines are available. In addition to the absolute visibility and closure phase, a spectro-interferometric instrument like AMBER produces differential quantities such as the differential visibility $V_{\mathrm{diff}}(\lambda)=V_{*}(\lambda)/V_{*}(\lambda_{\mathrm{r}})$ and the differential phase $\phi_{\mathrm{diff}}(\lambda)=\phi_{*}(\lambda)-\phi_{*}(\lambda_{\mathrm{r}})$, where $\lambda_{\mathrm{r}}$ is the wavelength of a reference channel, for example in the continuum near the emission line. As the differential measurements use an internal reference, they are much more accurate than the absolute measurements and allow to measure the small effects produced by quite unresolved sources.

 %The accuracy of the “self-calibrated” quantities, $V_{\mathrm{diff}}(\lambda)$ , $\phi_{\mathrm{diff}}(\lambda)$ and $\Psi(\lambda)$ is strongly dominated by the fundamental noise limits set by the photon noise, the detector noise and the thermal background photon noise, at least for MR observations over a small wavelength range.
 
 \subsubsection{Differential interferometry of non-resolved sources}

A non-resolved source has a global angular size $\Lambda$ that is smaller than the interferometer resolution limit $\lambda/B$. In Eq. \ref{eq:OI}, this implies that $O(\mathbfit{r},\lambda)$ is different from 0 only for values of $\mathbfit{r}<\lambda/\mathbfit{B}=1/\mathbfit{u}$, i.e. the integral in Eq. \ref{eq:OI} can be limited to values $\mathbfit{u} \cdot \mathbfit{r}<1$. \citet{1989dli..conf..249P} shows that the interferometric phase for such a source is given by
 \begin{equation}
 \phi_{\mathrm{*}}(\lambda, \lambda_{\mathrm{r}})=-2\pi \mathbfit{u} \cdot \left[\boldsymbol{\epsilon}(\lambda) -\boldsymbol{\epsilon}(\lambda_{\mathrm{r}}) \right],
 \end{equation}
 where the quantity
 \begin{equation}
  \boldsymbol{\epsilon}(\lambda)=\frac{\int\int{\mathbfit{r}O(\mathbfit{r}, \lambda)\, \mathrm{d^2}\mathbfit{r}}} {\int\int{O(\mathbfit{r}, \lambda)\, \mathrm{d^2}\mathbfit{r}}}
  \label{eq:photo}
 \end{equation}
 is the photocenter of the source. This result has been obtained from a first order limited development of the complex visibility $\tilde{O}(\mathbfit{u}, \lambda)$. Extending this development to higher terms shows \citep{2014SPIE.9146E..0R} that the source visibility $V_{*}$ is given by
 \begin{equation}
 V_{\mathrm{*}}=1-\alpha^2 \quad \text{where $\alpha = \frac{\pi \Lambda}{\sqrt{2}\left(\frac{\lambda}{B}\right)}$},
 \label{eq:vc}
 \end{equation}
 and the closure phase decreases as $\alpha^3$. The differential phase that decreases only proportionally to the source size has the highest super-resolution power, while the closure phase is accessible only on resolved sources. In section \ref{sec:present and future}, we discuss the accuracy limits on the differential visibility and phase for a list of AGNs, but let us first show how these measurements can be used. As a guideline for the reader we remind that for non-resolved sources, the differential phase gives the difference in position between the photocenter of the object in the reference channel and this in the $\lambda$ channel while the differential visibility gives their difference in size.

\section{Geometric and Kinematic model of the BLR}\label{sec:model}
 We developed a geometric and kinematic model of the broad line region surrounding a SMBH, to predict all interferometric and RM observables. This model is quite similar to the ones used when interpreting RM observations, for example by \citet{2012ApJ...754...49P}. Line profile modeling strongly suggests that the BLR is made of a very large number of clouds with individual sizes negligible with regard to the BLR size \citep{1997MNRAS.288.1015A,1998MNRAS.297..990A}. The covering factor of the BLR clouds is of the order of 10$\%$ \citep{2006A&A...456...75C,2012MNRAS.426.3086G} and most BLR clouds are seen directly by the observer if they are not masked by the dust torus. These basic hypothesis underlie RM and are well supported by its success.

%Since the BLR clouds need to absorb much of the UV and X-ray continuum to produce the large width of emission line seen in many Sy 1 objects hence it should have large opening angle. \citet{2006A&A...456...75C} suggest that $H/R$ should be larger than 0.1 and BLR disk could be a flared disk where scale $H$ increases more rapidly than linear with $R$. Bayesian Modeling of Reverberation Mapping Data of Mrk 50 by \citet{2012ApJ...754...49P} and Arp 151 by \citet{2011ApJ...733L..33B} found that the BLR has substantial scale height. This idea is further supported with the study of low-redshift quasars sample by \citet{2008MNRAS.387.1237D} which indicates an inflated disc BLR where C IV line is emitted in a flat disc while $\mathrm{H}\beta$ line originates in a geometrically thick region of BLR. Large quasar spectroscopic survey indicates C IV and Mg II have small dispersion in the distribution of line widths which implies that BLR can not be flat disc \citep{2008MNRAS.390.1413F, 2010MNRAS.409..591F}. Various authors proposed BLR model where line emitting gas spans from the outer accretion disc to the inner dust torus and the scale height increases with radial distance \citep{2012MNRAS.426.3086G,2009NewAR..53..140G}. The above evidences strongly suggest the BLR disk can not be thin, flat structure and hence we need to consider other possibilities.

%\subsection{Standard model setup}
\begin{figure}
\centering
\resizebox{9cm}{10cm}{\includegraphics{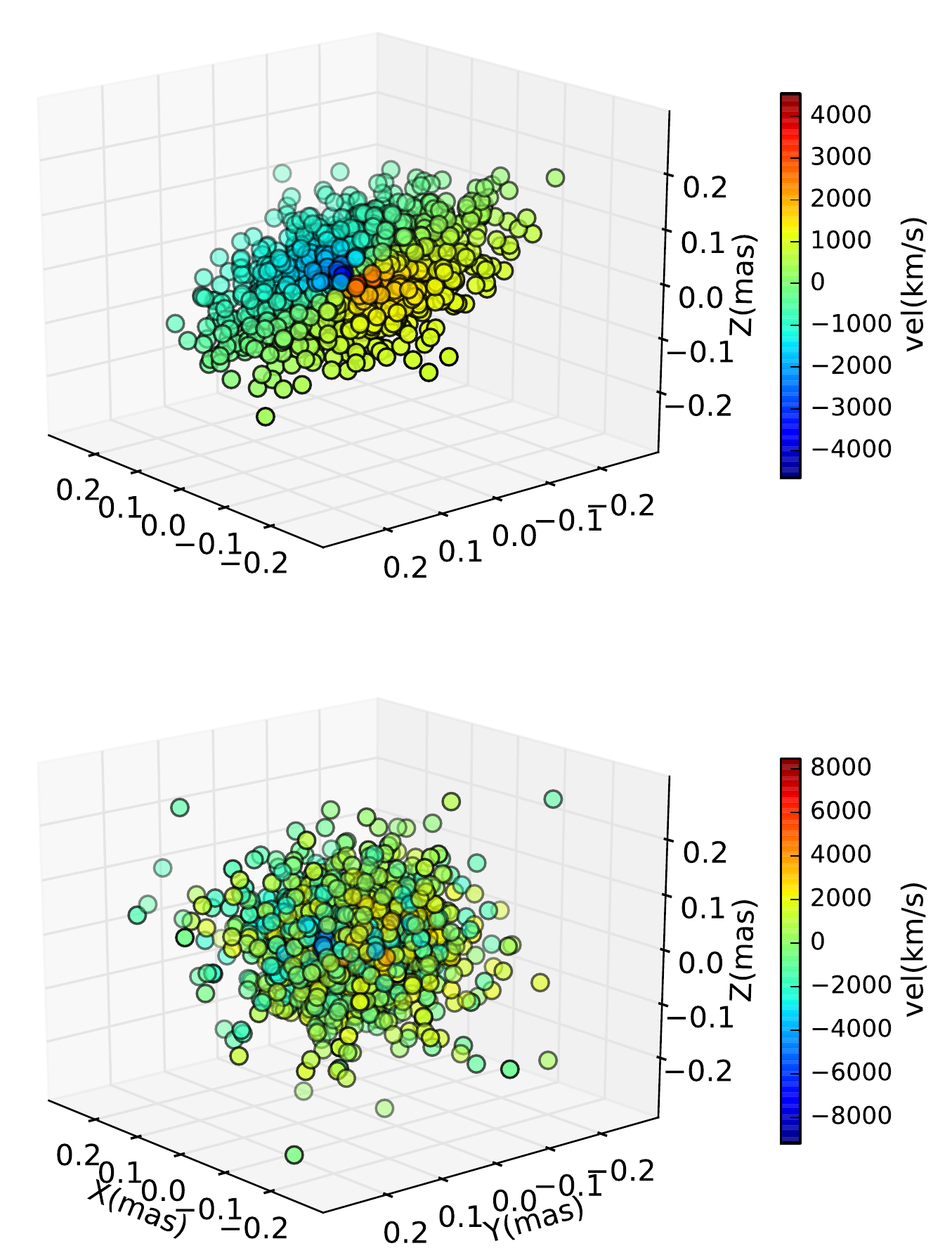}}
\caption{Cloud distribution with the velocity (km/s) in color code. $M_{\mathrm{bh}}=1e8M_{\mathrm{sun}}$, $i=30\degree$, $R_{\mathrm{in}}=1000R_g$, $\sigma_{\mathrm{blr}}=0.1$ mas with flat Keplerian disk geometry $\omega=0\degree$ (upper plot) and spherical geometry $\omega=90\degree$ (lower plot). }\label{Fig:dist3D}
\end{figure}

We start with a 3D distribution of clouds, considered as point sources. Each cloud is defined by its spherical coordinates $r$, $\theta$, $\phi$ in a reference system local to the BLR, with z as a symmetry axis. The coordinates are generated randomly with constraints of the desired BLR geometry.

The radius $r$ must respect the radial distribution of clouds $R_{d}(r)$ that is defined here as a normal distribution of standard deviation $\sigma_{\mathrm{blr}}$ limited by an inner radius $R_{\mathrm{in}}$ below which we assume the gas to be fully ionized and unable to contribute to a low ionization line.
\begin{equation}
P(l<r<l+dl)=\frac{1}{\sigma_{\mathrm{blr}}\sqrt{(2\pi)}} \mathrm{exp}\left(-\frac{l^2}{2\sigma_{\mathrm{blr}}^2}\right) \quad \text{for $ r \ge R_{\mathrm{in}} $}
\end{equation}

 %====================gridsize=======================

% {\begin{figure}
% \centering
% \resizebox{7cm}{15cm}{\includegraphics{\mypath/gridSIZE_SMALL.pdf}}
% \caption{Interferometric differential visibility(upper panel), phase (middle panel) and response function (lower panel) for different BLR spatial extension $\sigma_{\mathrm{blr}}$ and $\omega=0\degree$}\label{Fig:gridSIZE}
% \end{figure} 
 
  The declination angle $\theta$ is randomly selected to have -$\omega\le \theta \le \omega$, where $0\le \omega\le 90\degree$ is the “opening angle” of the BLR. For $\omega\simeq 0$ we have a flat disk. Many authors \citep{2006A&A...456...75C,2008MNRAS.387.1237D,2008MNRAS.390.1413F} have shown that a proper modeling of line profile implies a thick BLR that we choose to represent here using this opening angle as in \citet{2012ApJ...754...49P}. The azimuthal angle is randomly selected to have $0\le \phi \le 2\pi$. These spherical coordinates $r$, $\theta$, $\phi$ are used to define the tangential and radial velocity of the cloud, as discussed below. To maintain the axis symmetry of the cloud distribution an additional random rotation around z is added to each cloud position and velocity vector.

%The cloud is supposed to be on a circle of radius $r$ (a circular orbit if we are considering only orbital velocity), contained in the plan defined by the declination $\theta$, which is then randomly selected to have -$\omega\le \theta \le \omega$, where $0\le \omega\le 90\degree$ is the “opening angle” of the BLR. For $\omega\simeq 0$ we have a flat disc. Many authors \citep{2006A&A...456...75C,2008MNRAS.387.1237D,2008MNRAS.390.1413F} have shown that a proper modeling of line profile implies a thick BLR that we choose to represent using this opening angle as \citet{2012ApJ...754...49P}. The cloud can be anywhere on that circle, that is represented by a random azimuthal angle $0\le \phi \le 2\pi$. Since no orbital plane of declination must be privileged, an additional random rotation around the z-axis is introduced to restore the axisymmetry of the cloud distribution.
 
 From $r$, $\theta$, $\phi$ spherical referential we can define the cloud velocity vector considering several possible components:
 \begin{description}
 \item[a)] An orbital component, tangential to the circle: 
 \begin{equation}
 V_{\mathrm{orb}}  =V_{a}(\frac{R_{\mathrm{in}}}{r})^{\beta},
 \end{equation}
 where $\beta$ defines different rotation velocity laws \citep{1996A&A...311..945S}. For a Keplerian motion, $\beta =0.5$ and $V_{a}=\sqrt{\frac{GM_{\mathrm{bh}}}{R_{\mathrm{in}}}}$. 
 
 \item[b)] A radial component (inflow or outflow): 
 \begin{equation}
 V_{\mathrm{rad}} = V_{c}(\frac{R_{\mathrm{in}}}{r})^\gamma,
 \end{equation}
 where $\gamma$ is the power law index of this radial velocity. Freefall corresponds to $\gamma=0.5$ and $V_{c}=\sqrt{\frac{2GM_{\mathrm{bh}}}{R_{\mathrm{in}}}}$. $\gamma=-1$ is an outflow case with outflow velocity amplitude $V_{c}$ set at the inner radius $R_{\mathrm{in}}$ of the BLR \citep{1991ApJ...379..586W}.
 \end{description}
 
 The composition of these orbital and radial velocity laws constitutes the global velocity field of the BLR. We have also considered a local macroturbulent velocity component $V_{\mathrm{turb}}$ with random orientation. Several authors \citep{2006A&A...456...75C,2012MNRAS.426.3086G} relate the amplitude $V_{\mathrm{turb}}$ to the thickness $H(r)$ of the BLR at the radius $r$ 
 \begin{equation}
 |V_{\mathrm{turb}}|=V_{\mathrm{orb}}\,P_{\mathrm{turb}}\, H(r).
 \label{eq:macro}
 \end{equation}
In our model $H(r)=r\,\mathrm{sin}\,\omega$. The multiplicative parameter $P_{\mathrm{turb}}$ tunes the amplitude of the turbulence velocity. The macroturbulence is zero both for flat disks ($\omega=0$) and for $P_{\mathrm{turb}}=0$.

 To obtain the geometrical distribution with respect to the observer, in its X,Y,Z coordinate system, we introduce a global rotation about y-axis of angle $i$ ($i$=0 is for “face on” object, and for Sy1 the typically value of $i$ is less than $40\degree$) and then a global rotation around the Z-axis to introduce the position angle $\Theta$. In the following we consider that $\Theta$ is known, as it can be deduced either from the jet position angle, from polarization measurements or from broadband OI observation. It is also possible to introduce this parameter in the model fit as soon as we have several baseline orientations.

The cloud apparent brightness can be affected by a geometrical effect related to its optical thickness and to its position, named “anisotropy” by several authors \citep{1994MNRAS.268..845O,2012MNRAS.426.3086G,2013arXiv1311.6475P,2014arXiv1407.2941P}. If the cloud is optically thick, then the observer sees only the fraction of its surface facing him. If the cloud is optically thin, then all points of the cloud contribute to its intensity in all directions. This effect, similar to a “moon phase”, is described as
\begin{equation}
I(\phi)=(1-F_{\mathrm{anis}}\,\mathrm{cos}\,\phi\,\mathrm{sin}\,i) ,
\label{eq:anis}
\end{equation} 
 where the anisotropy factor $F_{\mathrm{anis}}$ goes from 0 for optically thin clouds to 1 for optically thick clouds. In the present stage of development, our model includes only this simple description of anisotropic cloud emission as well as the simple anisotropy of a skewed torus inner rim described in the section \ref{sec:signal_estimation}. A more detailed analysis of the anisotropy of the BLR emission as well as of its dust surroundings has to be undertaken in a future paper through a radiative transfer modeling using for example the photo-ionizing code Cloudy \citep{1998PASP..110..761F,2013RMxAA..49..137F} that constrains the cloud opacity and provides the radial distribution of its emission.

Each cloud is emitting a line with profile $L_{\mathrm{XYZ}} (\lambda)$ depending from the local physical conditions and hence from the cloud position. This profile is convolved by the instrument spectral PSF $P_{I} (\lambda)$. If we observe at relatively low spectral resolution, from 200 to 1500, we can consider that $P_{I} (\lambda)$ is much broader than $L_{\mathrm{XYZ}} (\lambda)$ and, as a first approximation, we can consider that the line shape details are lost in the convolution. Thus, in the current version of our model all clouds are emitting the same line profile $L(\lambda)$, but for its intensity that depend on $r$ and can be included in the radial intensity distribution $R_{d}(r)$. This gives $L_{\mathrm{XYZ}} (\lambda)\, P_{I} (\lambda) \simeq R_{d}(r)\, L(\lambda)$. We choose to represent the local line profile $L(\lambda)$ by a Gaussian function centered at the emission line wavelength $\lambda_0$ and with standard deviation $\sigma_0$ that is one of the parameters of the model: $L(\lambda)=\mathcal{N}(\lambda-\lambda_0,\sigma_0)$.
%--------------------------------------------------------------------
\begin{center}
 \begin{table}
 \caption{Model parameters and the observables.}
 \begin{minipage}{0.5\textwidth}
 	\centering {
 	\small\addtolength{\tabcolsep}{3pt}
     \begin{tabular}{ l l l}   
     \hline
     \hline
     Parameter &  Symbol & Ref. value\footnote{if not stated.}\\
     \hline
     BH mass    & $M_{\mathrm{bh}}$ & $1e8\,M_{\mathrm{sun}}$ \\
     BLR inner radius & $R_{\mathrm{in}}$ & $200R_{g}$ \\
     BLR width  & $\sigma_{\mathrm{blr}}$ & 0.4 mas  \\
     Inclination & $i$ & $30\degree$ \\
     Opening angle & $\omega$ & 0$\degree$  \\
     Rest line width & $\sigma_0$ & 85 km/s  \\
     Macroturbulence & $P_{\mathrm{turb}}$ & 0 \\
     Anisotropy  & $F_{\mathrm{anis}}$  & 0     \\
     Continuum size & $R_{\mathrm{rim}}$ & 0.25 mas \\
     Disk position angle & $\Theta$ & 90$\degree$ \\
     Object Redshift     & z     & 0.02 \\
     Emission line flux & $F$  & 0.6   \\                
     \hline
     Measure & Symbol     & Observing Technique  \\
     \hline
     Absolute visibility & $V_{\mathrm{abs}}(\lambda)$ & OI  \\
     Differential visibility & $V_{\mathrm{diff}}(\lambda)$ & OI \\
     Differential phase & $\phi_{\mathrm{diff}}(\lambda)$   & OI  \\
     Spectrum &  $s(\lambda)$  & RM or OI \\
     2D Response function & $\psi(v, \tau)$  & RM  \\
     1D Response function &  $\psi(\tau)$ & RM \\
     \hline
 	\end{tabular} }
 	\end{minipage}
 \end{table}
 \end{center}
 %--------------------------------------------------------------
 \begin{figure*}
     \centering
     \setlength{\unitlength}{1cm}
     \begin{picture}(18, 10)
     \put(0.2, 6){\includegraphics[width=17.5cm, height=4cm]{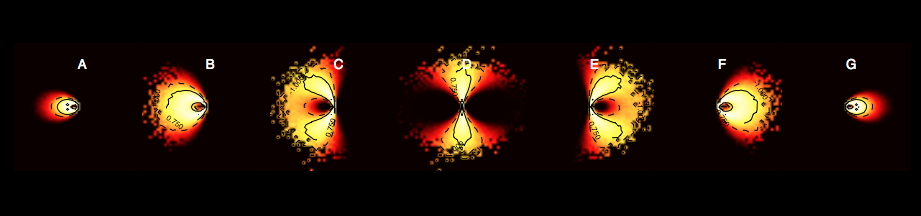}}  %line_SMALL.png
     %\resizebox{18cm}{4cm}{
     %\put(1, 0){\includegraphics[width=16.0cm, height=5cm]{\mypath/Photocenter_Spect.pdf}}
     %\put(1, 0){\includegraphics[width=16.0cm, height=6cm]{\mypath/Spect_photo.pdf}}
     \put(1, 0){\includegraphics[width=16.0cm, height=6cm]{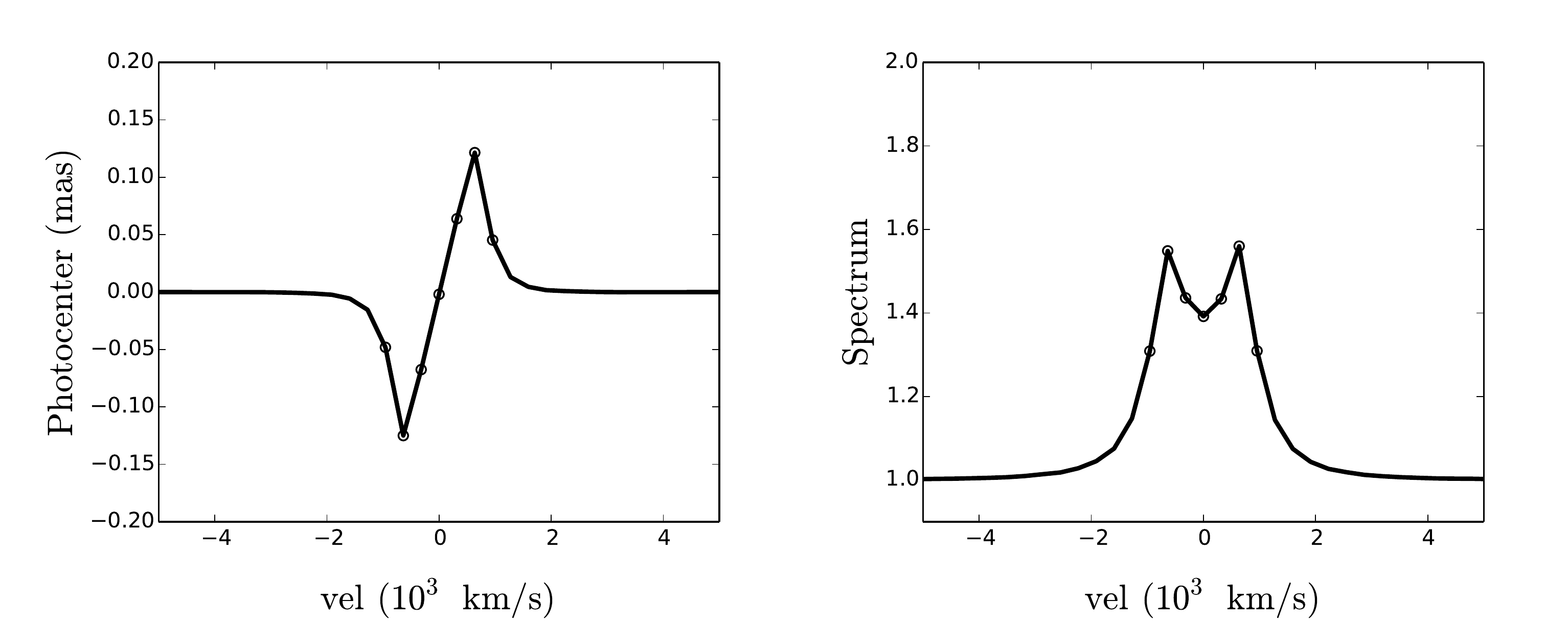}}
     \put(4.5, 2.5){A}
     \put(5, 1.5){B}
     \put(5.4, 2.3){C}
     \put(5.6, 3){D}
     \put(5.3, 3.8){E}
     \put(5.8, 4.7){F}
     \put(6.2, 3.6){G}
     
     \put(12.4, 2.5){A}
     \put(12.8, 3.8){B}
     \put(12.9, 2.8){C}
     \put(13.2, 2.6){D}
     \put(13.4, 2.8){E}
     \put(13.6, 3.8){F}
     \put(13.95, 2.5){G}     
     \end{picture}
      \caption{Line intensity map (upper panel) across the emission line for a flat Keplerian  disk geometry for seven different spectral channels (marked with the letter). Photocenter displacement perpendicular to the rotation axis (lower-left panel) and emission line profile (lower-right panel) for different spectral channels with resolution $R=1500$ is plotted. This model is computed considering $\sigma_{\mathrm{blr}}=0.4$ and $i=30\degree$. }
       \label{Fig:line-Spect}   
 \end{figure*}
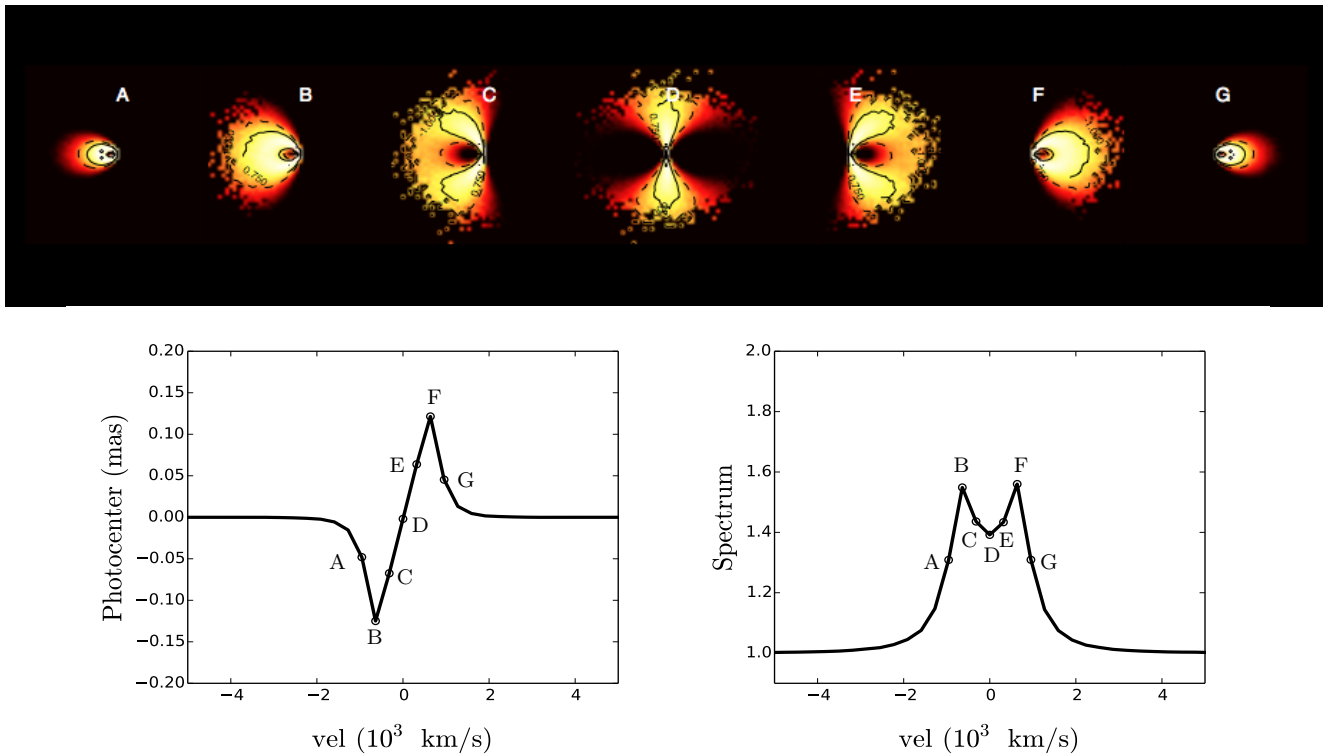
 %-------------------------------------------------------------------
%\footnote{\bf Relativistic effects, such as relativistic Doppler shift and gravitational redshift can significantly change the shape of the emission line profile if the emission line clouds are located close to the central source \citep{2012MNRAS.426.3086G,2014MNRAS.445.3055P}. However, in this paper we are mainly focusing on the low-ionization hydrogen lines such as H, Paschen and Bracket series, which are produced far enough from the central source for the relativistic effect to be negligible.} 

For each cloud, the line is Doppler shifted by the projected velocity component $V_{z} (X,Y,Z)$ and  the contribution of each cloud to the BLR intensity is
{\small
\begin{equation}
I_{\mathrm{blr}} (X,Y,Z,\lambda)=R_d (X,Y,Z)\,\mathcal{N}\left[\lambda-\lambda_0 \left(1+ \frac{V_{z} (X,Y,Z)}{c} \right),\sigma_0\right].
\end{equation}} 
Since we are focusing on low-ionization hydrogen lines such as Paschen and Bracket series that are produced far from the central source, the relativistic effects such as relativistic Doppler shift and gravitational redshift are negligible and hence we do not take them into account.
%\citet{2012MNRAS.426.3086G} and \citet{2014MNRAS.445.3055P} shows that these effects can significantly change the shape of emission line profiles if the emission line clouds are located close the central source.}}

The BLR intensity seen by the observer $I_{\mathrm{blr}} (X,Y,\lambda)$ is obtained by adding the contribution $I_{\mathrm{blr}} (X,Y,Z,\lambda)$ of all clouds located in a (X, X+dX, Y, Y+dY) box and by normalizing it with its maximum amplitude. 
Fig. \ref{Fig:dist3D} represents two distributions of cloud. The upper panel shows a flat Keplerian disk ($\omega=0$, $\beta=0.5$, $V_{\mathrm{rad}}$=0, $i=30\degree$, $R_{\mathrm{in}}=1000 R_{g}$, where $R_{g} \equiv GM_{\mathrm{bh}}/c^2$). The lower panel shows a spherical distribution of Keplerian orbits ($\omega=90\degree$). The colors represent the velocity in the direction of the observer. Note that the velocity range in the spherical case is twice larger than the flat Keplerian case with the same central mass and BLR size.
 %=====================plot line image================================
\\
{\bf Continuum model:} In the $K$-band, the continuum emission is strongly dominated by the hot dust near the sublimation radius $R_{\mathrm{rim}}$ \citep{2007A&A...476..713K, 2009A&A...493L..57K}. As this structure remains unresolved both for the VLTI and the KI, we have access only to its equivalent radius. We choose to represent it by a narrow ring whose radius ($R_{\mathrm{rim}}$) will give the right visibility observed in the continuum, when such a measurement is available, or $ \propto L^{0.5}$ with a proportionality constant deduced from \citet{2006ApJ...639...46S}. For most of this paper, we consider that the ring is uniform and we do not introduce any skewing related to the inclination, although such a function is easy to introduce in a parametric form. A skewing of the continuum image, with a “face on” side brighter than the “back on” side will introduce a measurable phase effect that is briefly discussed in section \ref{sec:signal_estimation}. The continuum brightness distribution $I_{\mathrm{con}} (X,Y)$ is normalized to have $\int{\int{I_{\mathrm{con}} (X,Y) \mathrm{d}X \mathrm{d}Y}}=1$.
\\
{\bf Model parameters and its observables:} The global intensity is obtained by adding the BLR and continuum intensities
%{\small
\begin{equation}
I(X, Y, \lambda) =I_{\mathrm{con}}(X, Y) +F\,I_{\mathrm{blr}}(X, Y, \lambda),
\end{equation}
where $F$ is the maximum emission line flux, for a measured spectrum $S_{M}(\lambda_c)$=1 in the continuum. 

This $F$ factor is a characteristic of the emission line chosen for the interferometric observations. The equations show that all differential interferometry signals are proportional to $F/(1+F)$ but their shape is unaffected. This means that the value $F$ matters only when we are discussing the signal-to-noise ratio and the feasibility of interferometric observations. In section \ref{sec:Obs_signatures}, where we discuss the parameters signatures, we use $F=0.6$, which is the value we observed on 3C273 for $\mathrm{Pa}\, \alpha$. It is also close to the $F=0.69$ mean $\mathrm{Pa}\, \alpha$ value that can be found in \citet{2008ApJS..174..282L}. In the feasibility section \ref{sec:present and future} we use the best emission line in the $K$-band set by the redshift of each target. This line is chosen among $\mathrm{Pa}\, \alpha$, $\mathrm{Pa}\, \beta$, $\mathrm{Pa}\, \gamma$ or $\mathrm{Br}\, \alpha$ and the line strength elements given in \citet{2008ApJS..174..282L} are taken in to account in section \ref{sec:present and future}.

A Fourier transform of the intensity distribution $I(X,Y,\lambda)$ yields the complex visibility, its modulus and phase, with the subsequent absolute and differential visibility and differential phase as mentioned in section \ref{sec:interferometery}.

The time delay, between the continuum and the emission line echo, is a function of the corresponding cloud position and defined by 
 \begin{equation}
 \tau(r, Z)=\frac{r-Z}{c}. 
 \end{equation} 
 We compute normalized histogram of time delays $\tau$ and velocities $V_z$  in the observer direction to obtain the 2D echo diagram $\Psi(v, \tau)$. The integration of $\Psi(v, \tau)$ over time gives the 1D response function $\Psi(\tau)$, whose centroid is the equivalent time lag $\tau_{\mathrm{cent}}$, and the integration over velocity gives the 1D line profile $\Psi(v)$ whose width yields the equivalent range $\Delta V$ of the global velocity field.

For a flat Keplerian disk model narrow-band line images are plotted in the upper panel of Fig. \ref{Fig:line-Spect} for different spectral channels. As we enter the line at maximum redshift, a line image appears in addition to the continuum, the photocenter shifting perpendicularly to the rotation axis. The photocenter shifts (lower-left panel) goes through an extrema around channels B and C, then cancels in channel D at the center of line, where the image is symmetric. The blue wing images (E to G) mirror the red wing and the photocenter is shifted in the opposite direction. The emission line profile (lower-right panel) shows a double peaked profile as expected for a thin Keplerian rotation. The line intensity shows maxima (B and F channels) related to the inclination and the equivalent outer edge of the BLR. We see that the images in channels B and F show maximum intensity and extension in the direction $\perp$ to the axis. This corresponds to local minima in the visibility in $\perp$ baseline. The maximum intensity and extension in the $\parallel$ direction is in channel D yields the local visibility minima in $\parallel$ baseline.

%As velocity goes from red to blue side, line starts appearing on the top of the continuum, the position of the photocenter changes toward South then become zero as the iso-velocity curves extends equally both directions and finally leaves the continuum as velocity becomes high. Emission line profile (lower panel of Fig. \ref{Fig:line-Spect}) shows a double peaked profile as expected for a thin Keplerian rotation. For a fixed inclination the spectrum shows a peak at  $\pm V_{z}(R_{\mathrm{out}})$ and wings extending to $\pm V_{z}(R_{\mathrm{in}})$.
%===========================================================
%=======================Model================================

%===========================================================

%####################################### Absolute and differential visbility #############################
%####################################### ----------------------------------- #############################

\section{Observable signatures of the model parameters}
\label{sec:Obs_signatures}
In this section we illustrate the effect of the main model parameters on the OI and RM observables. We have tried to analyze the parameters in an order that allows to partially separate their observable effects and isolate typical spectro-interferometric signatures. We will discuss first the measurement of the equivalent angular sizes of the BLR that is mostly related to the global amplitude of visibility measurements. Then we will examine how differential visibility and phase can solve the major BLR model ambiguities: the degeneracy between inclination ($i$), thickness ($\omega$) and the balance between local and global velocity field, that we chose to represent, in a first step, by the width $\sigma_0$ of the line profile including all local velocity effects. Finally, we will show how the components of the global velocity field can be separated by differential phase measurements and examine the signatures of other physical phenomena such as the clouds optical thickness and the macroturbulent component of the local velocity field.

\begin{figure}
 \centering
 \resizebox{8.6cm}{4cm}{\includegraphics{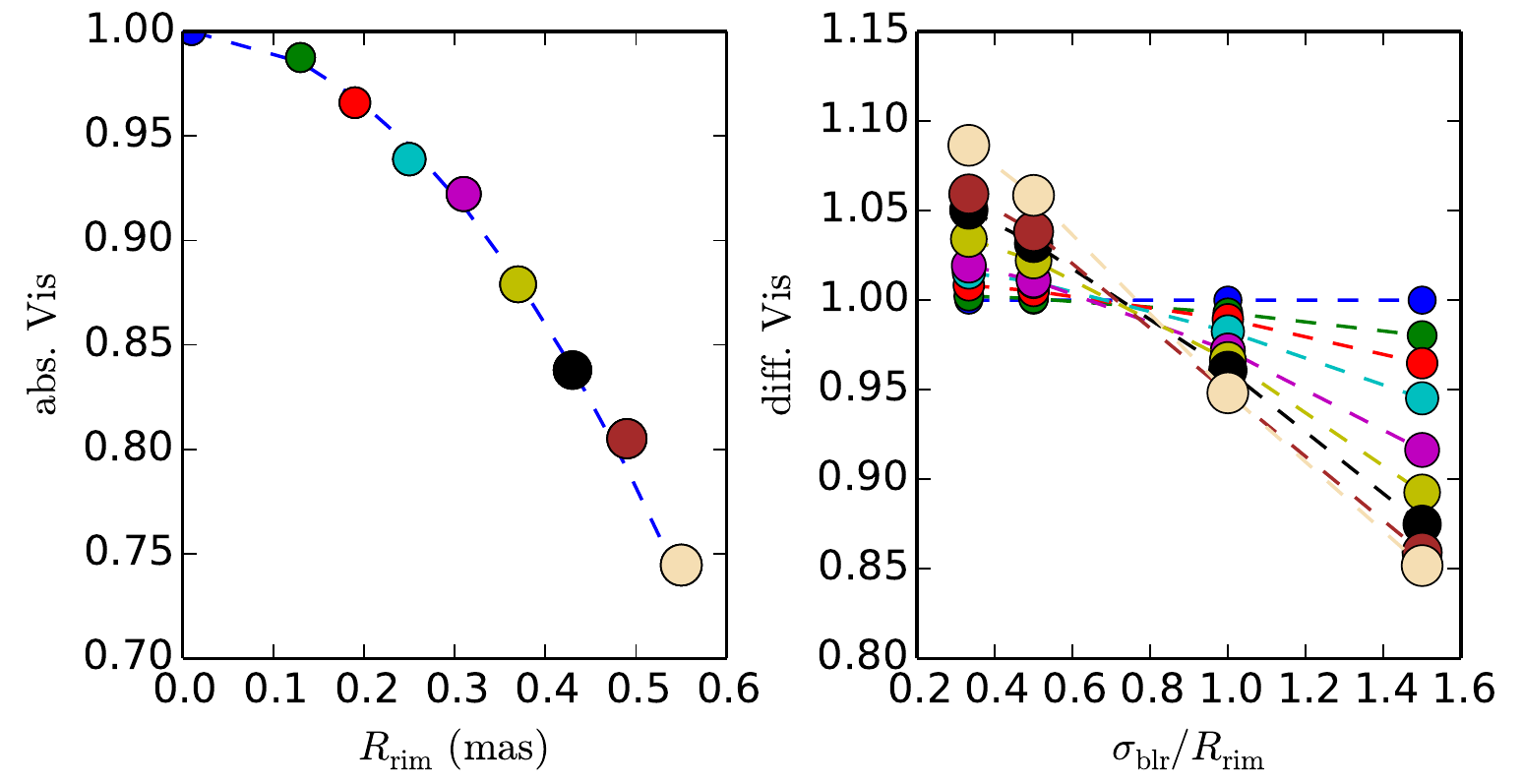}}  %or put abs_diff_vis_SMALL.pdf 
 
 %\resizebox{6.0cm}{12.0cm}{\includegraphics{\mypath/continuum_SMALL.pdf}}
 %\resizebox{6cm}{12cm}{\includegraphics{\mypath/gridSIZE_SMALL.pdf}}
 %\resizebox{6.0cm}{12cm}{\includegraphics{\mypath/abs_diff_vis_SMALL.pdf}}
 \caption{Absolute visibility in the continuum as a function of $R_{\mathrm{rim}}$ (left panel) and differential visibility in the line as a function of $\sigma_{\mathrm{blr}}/R_{\mathrm{rim}}$ (right panel) for 130 m baseline and parallel to the rotation axis. }\label{Fig:cont}
 \end{figure} 

\begin{figure}
\centering
\resizebox{8.6cm}{7cm}{\includegraphics{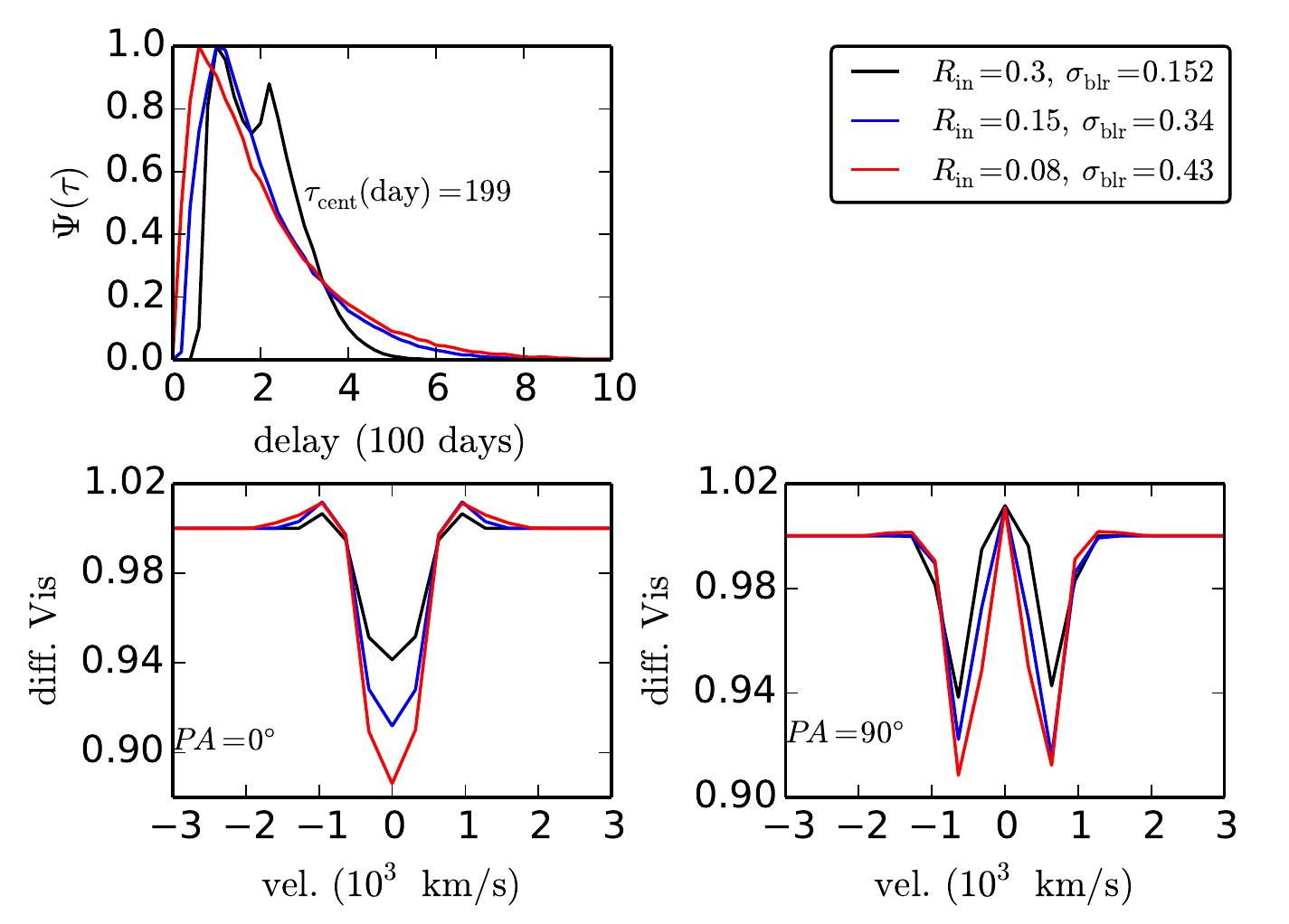}}
\caption{RM 1D response function (upper panel) and visibility in the $\parallel$ (lower-left panel) and the $\perp$ (lower-right panel) baselines for different BLR geometries that produce same $\tau_{\mathrm{cent}}$ but different visibilities.}
\label{Fig:SameSitau}
\end{figure}

\subsection{Interferometric BLR size}\label{sec:BLRsize}

An estimation of the angular size of the BLR is critical to constrain the $r_{\mathrm{blr}}-L$ and the $M_{\mathrm{bh}}-L$ relations. Combined with the RM linear size measurement, it can yield a direct distance measurement \citep{2002ApJ...581L..67E}. This angular size can be constrained by a broadband measurement of the absolute visibility in the continuum combined with a relatively low spectral resolution differential visibility measurement with only one measurement in the emission line. 

The left panel of Fig. \ref{Fig:cont} displays the absolute visibility as a function of $R_{\mathrm{rim}}$. With a typical visibility accuracy of current VLTI instruments $\sigma_{\mathrm{avis}}\simeq0.03$, we see that the smallest $R_{\mathrm{rim}}$ that can be estimated in the $K$-band with the VLTI baselines is $R_{\mathrm{rim}}\simeq 0.15$ mas. A fringe tracker, such as the one built-in in GRAVITY \citep{2008SPIE.7013E..69E}, should allow to reduce the absolute visibility error down to $\sigma_{\mathrm{avis}}\simeq 0.005$, making it possible to measure down to $R_{\mathrm{rim}}\lesssim 0.06$ mas.

The right panel in Fig. \ref{Fig:cont} displays the differential visibility as a function of the ratio of $\sigma_{\mathrm{blr}}/R_{\mathrm{rim}}$. This measurement exhibits a very good accuracy ($\sigma_{\mathrm{dvis}}\lesssim 0.001$) limited only by fundamental noises. It cannot yield the relative sizes $\sigma_{\mathrm{blr}}/R_{\mathrm{rim}}$ when $|V_{\mathrm{diff}}-1|<0.001$, if $R_{\mathrm{rim}}\lesssim 0.1$ mas at the VLTI i.e. $(\lambda/B)/R_{\mathrm{rim}}\gtrsim 35 $. This sets a limit for the super-resolution factor that can be expected from visibility measurements. The uncertainty on the angular size of the BLR will be dominated by the absolute visibility accuracy that is therefore a key specification for BLR size estimation. Note that Fig. \ref{Fig:cont} confirms the prediction of Eq. \ref{eq:vc} and sets the equivalence between the size parameters $R_{\mathrm{rim}}$ and $\sigma_{\mathrm{blr}}$: a flat Keplerian BLR model produces the same visibility than a thin ring when $\sigma_{\mathrm{blr}}/R_{\mathrm{rim}} \simeq 0.7$. If we have differential visibilities for two different baselines (with $(\lambda/B)/R_{\mathrm{rim}}\gtrsim 35 $ for the shortest baseline) we can obtain  $\sigma_{\mathrm{blr}}$ and $R_{\mathrm{rim}}$ without absolute visibility measurements, but the accuracy of this method has not been evaluated yet.
%################## Visbility and response function ##########################
%###################---------------- ---------------#########################

%===========================================================
%=======================Thick geometry======================
%
%\vspace*{25cm}
%\newpage

%Insert Figure Mass and f sensitivity to i and omega.
  \begin{figure}
  \centering
  \resizebox{8.6cm}{4cm}{\includegraphics{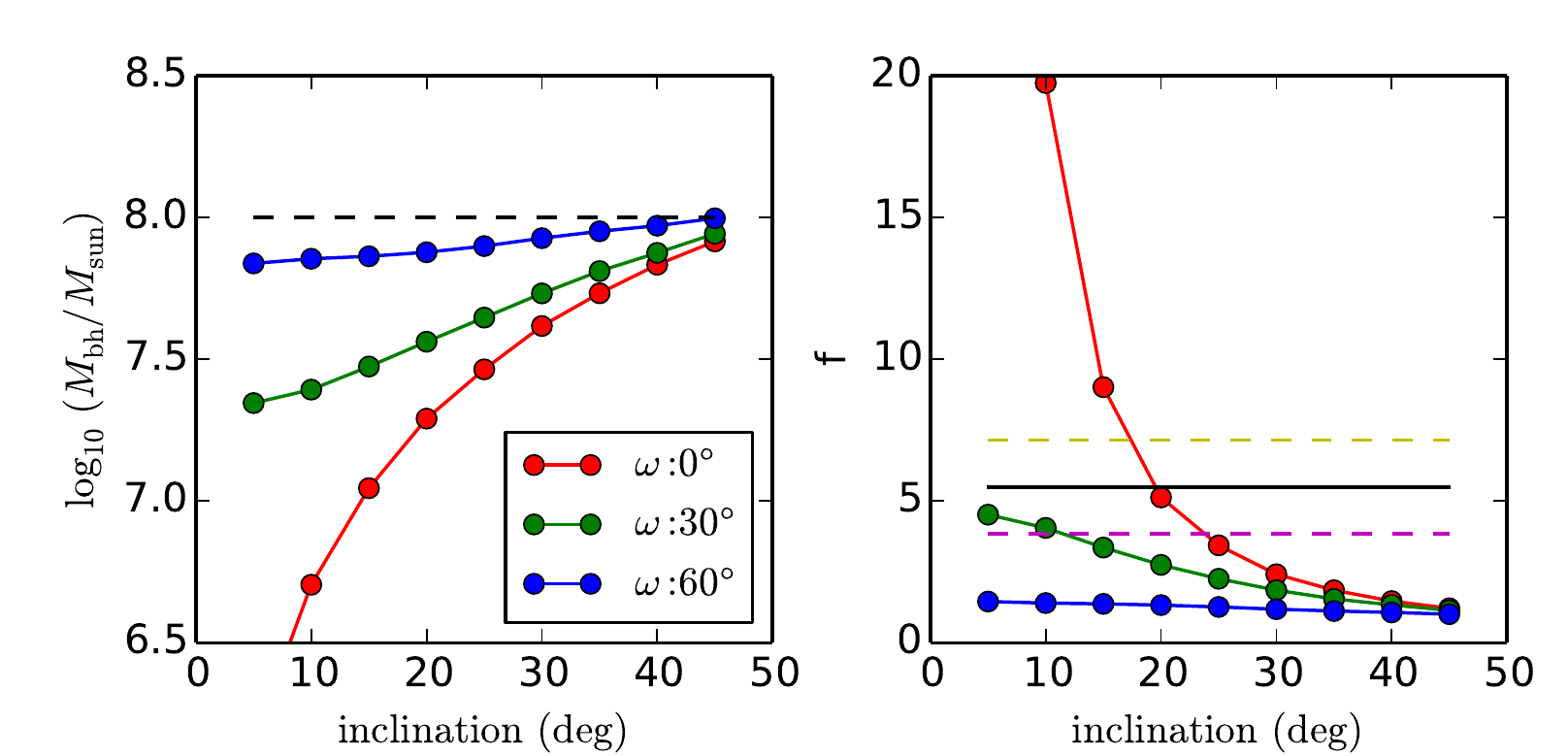}} %mass_ratio_SMALL.pdf
  \caption{$M_{\mathrm{bh}}$ (left panel) and scale factor (right panel) as a function of inclination for different opening angles $\omega=0\degree$ (red), $30\degree$ (green) and $60\degree$ (blue). The input mass of this simulation is $10^8 \, M_{\mathrm{sun}}$. We see that an error on $i$ or $\omega$ can result in a very large mass error.
  }
  \label{Fig:MassRatio}
  \end{figure}

\subsection{Interferometric and Reverberation Mapping BLR sizes}\label{sec:vis-res} 

The different parts of the source contribute to the interferometric and RM sizes with different weights. To illustrate this, we consider different flat geometries with different combinations of $R_{\mathrm{in}}$ and $\sigma_{\mathrm{blr}}$ that produce the same equivalent time lag $\tau_{\mathrm{cent}}$, from a hollow thin torus (large $R_{\mathrm{in}}$ and small $\sigma_{\mathrm{blr}}$, black line in Fig. \ref{Fig:SameSitau}) to an extended BLR with almost no central hole (small $R_{\mathrm{in}}$ and large $\sigma_{\mathrm{blr}}$, red line in Fig. \ref{Fig:SameSitau}). Fig. \ref{Fig:SameSitau} shows that these combinations produce very different visibilities. The peak of $\Psi(\tau)$ grows with $R_{\mathrm{in}}$ but the centroid $\tau_{\mathrm{cent}}$ remains constant. The overall shape of the differential visibility $V_{\mathrm{diff}}(\lambda)$ curves remains stable, respectively looking like a “v” and “w” for the $\parallel$ and the $\perp$ baselines. Their amplitude is almost proportional to $\sigma_{\mathrm{blr}}$. Measuring QSO distances from a combination of OI and RM observations requires a calibration of this effect that will also influence the size-luminosity relation. This will be the scope of next paper on the combination of OI with RM.

 \begin{figure*}
 \centering
 %\resizebox{8cm}{10cm}{\includegraphics{\mypath/gridSPECT.png}}
 \resizebox{8.5cm}{8.0cm}{\includegraphics{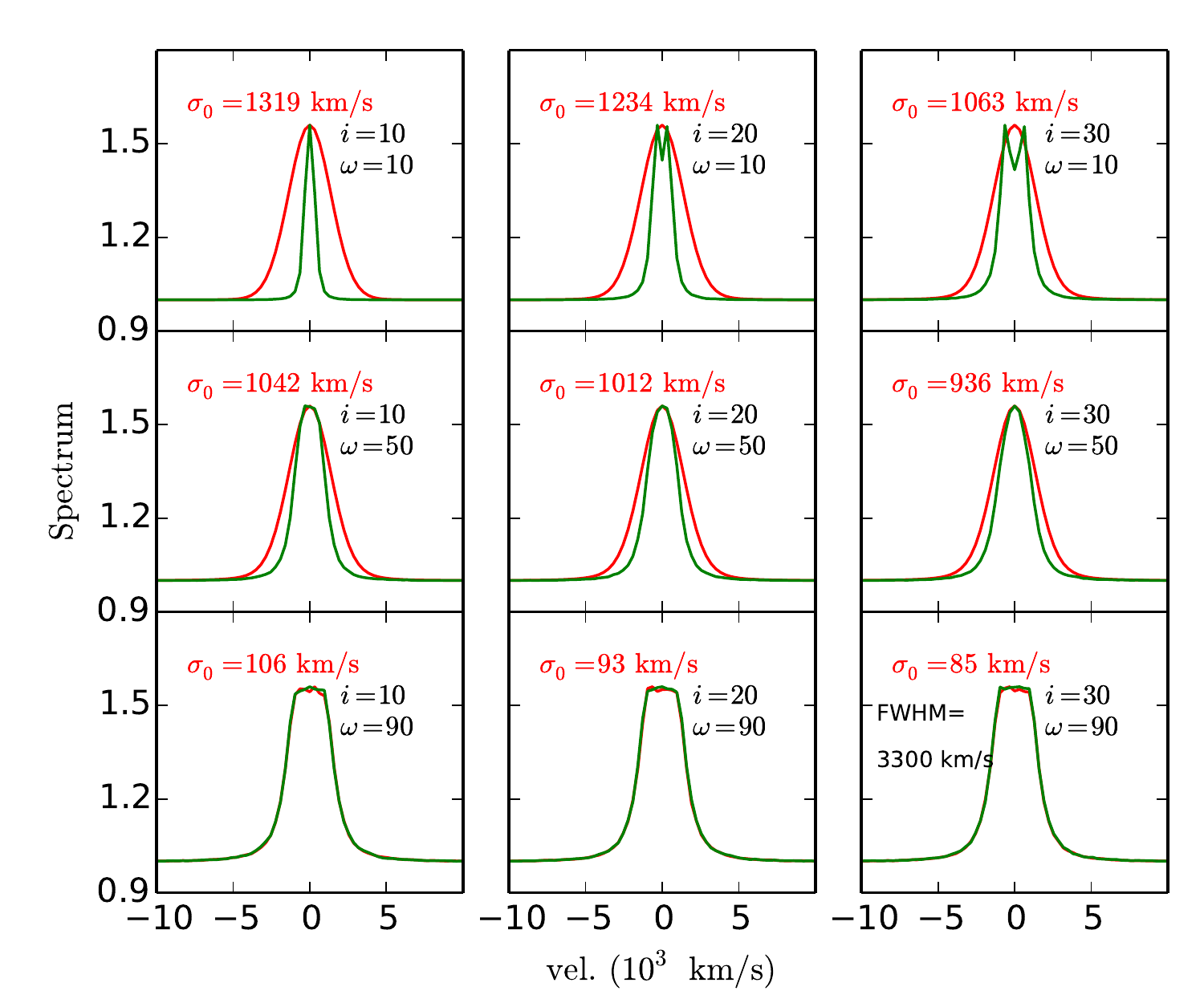}}
 \resizebox{8.5cm}{8cm}{\includegraphics{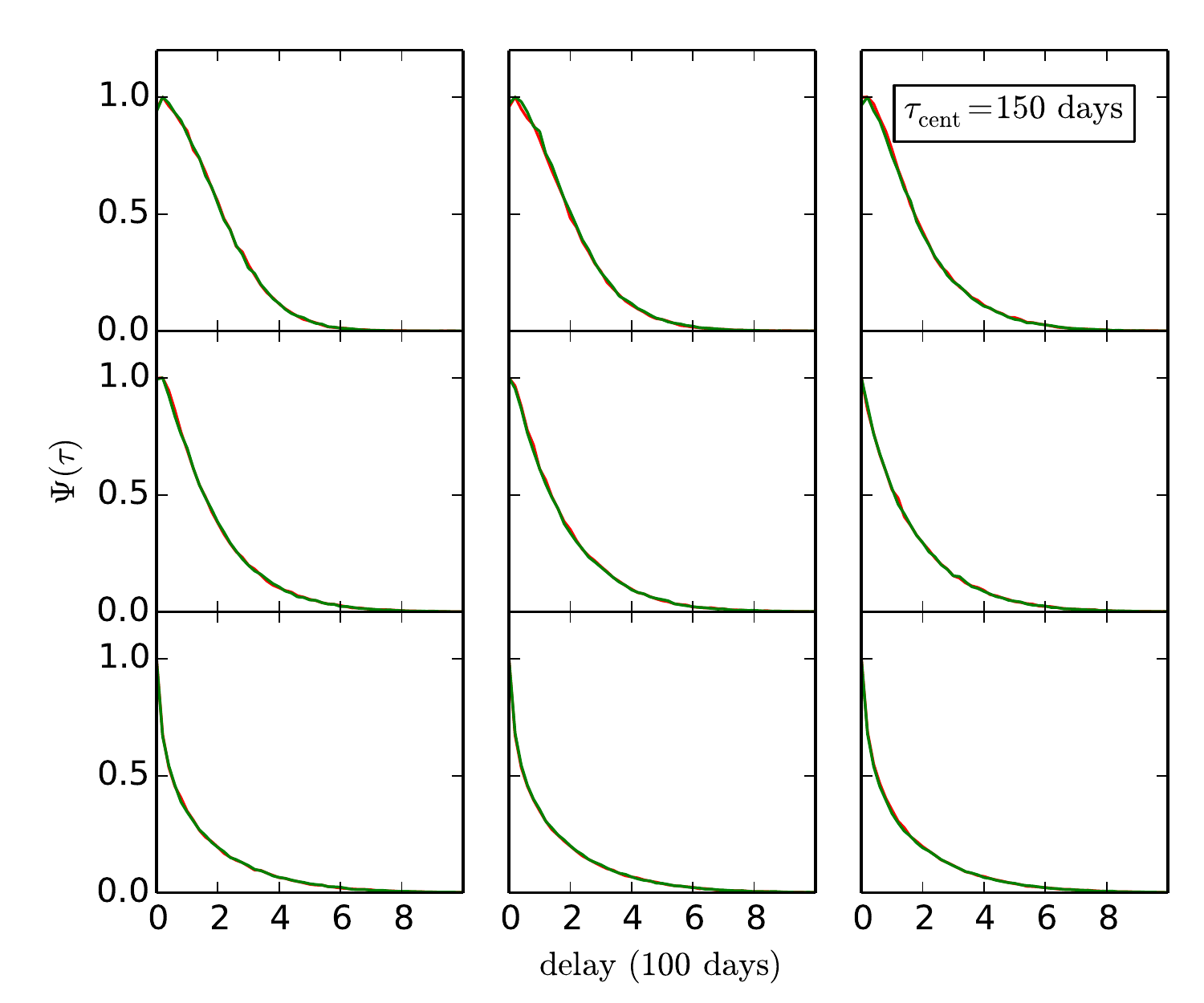}}
 \caption{Spectrum (left panel) and 1D response function $\Psi(\tau)$ (right panel) for different inclinations and opening angles. Green curves in each plot obtained with $\sigma_0$=85 km/s whereas the red curves are for different $\sigma_0$ values as mentioned in the left panel.}\label{Fig:gridSPECT-SITAU}
 \end{figure*}
 
 \begin{figure*}
  \centering{
  \resizebox{8.5cm}{8cm}{\includegraphics{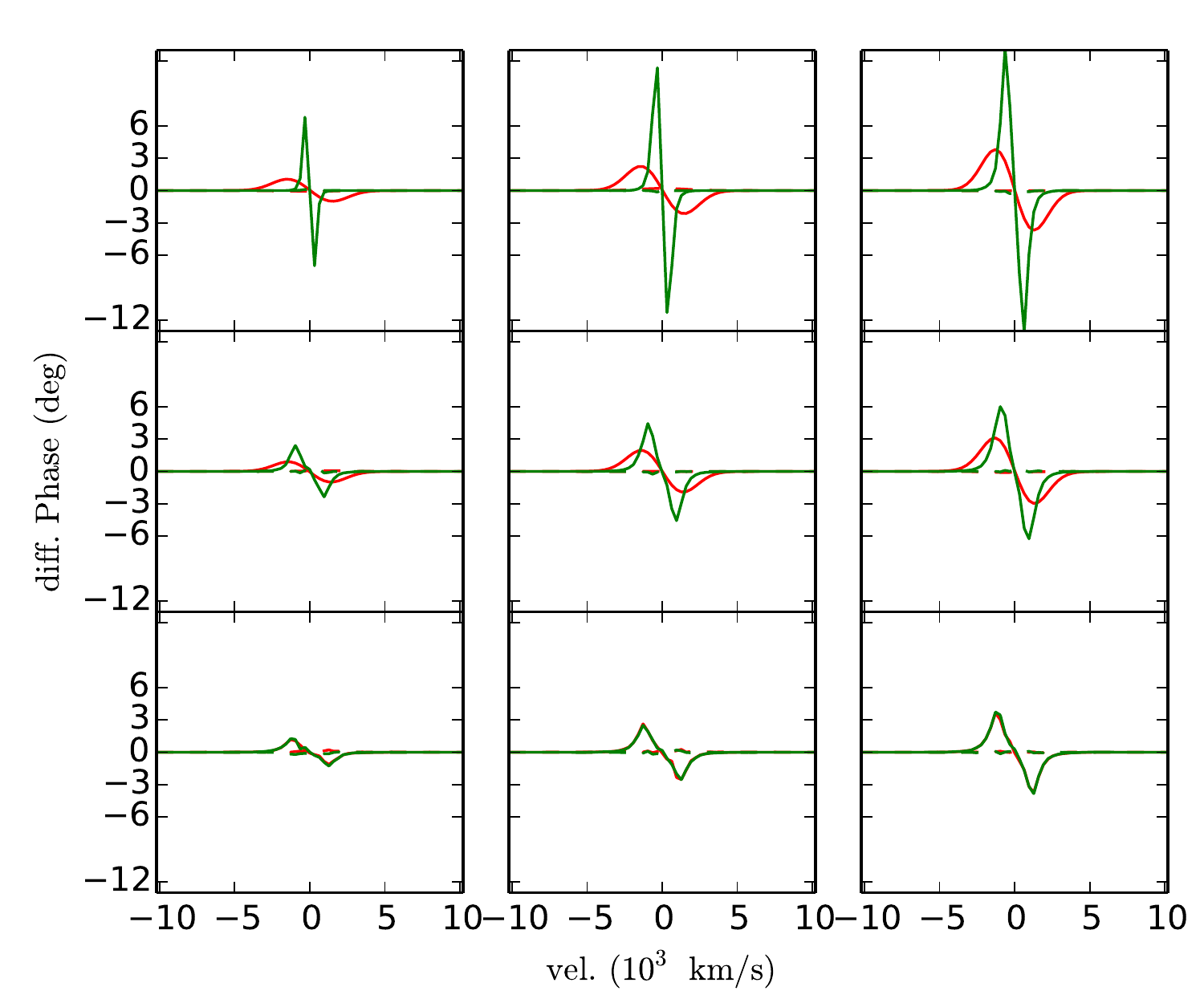}}
  \resizebox{8.5cm}{8cm}{\includegraphics{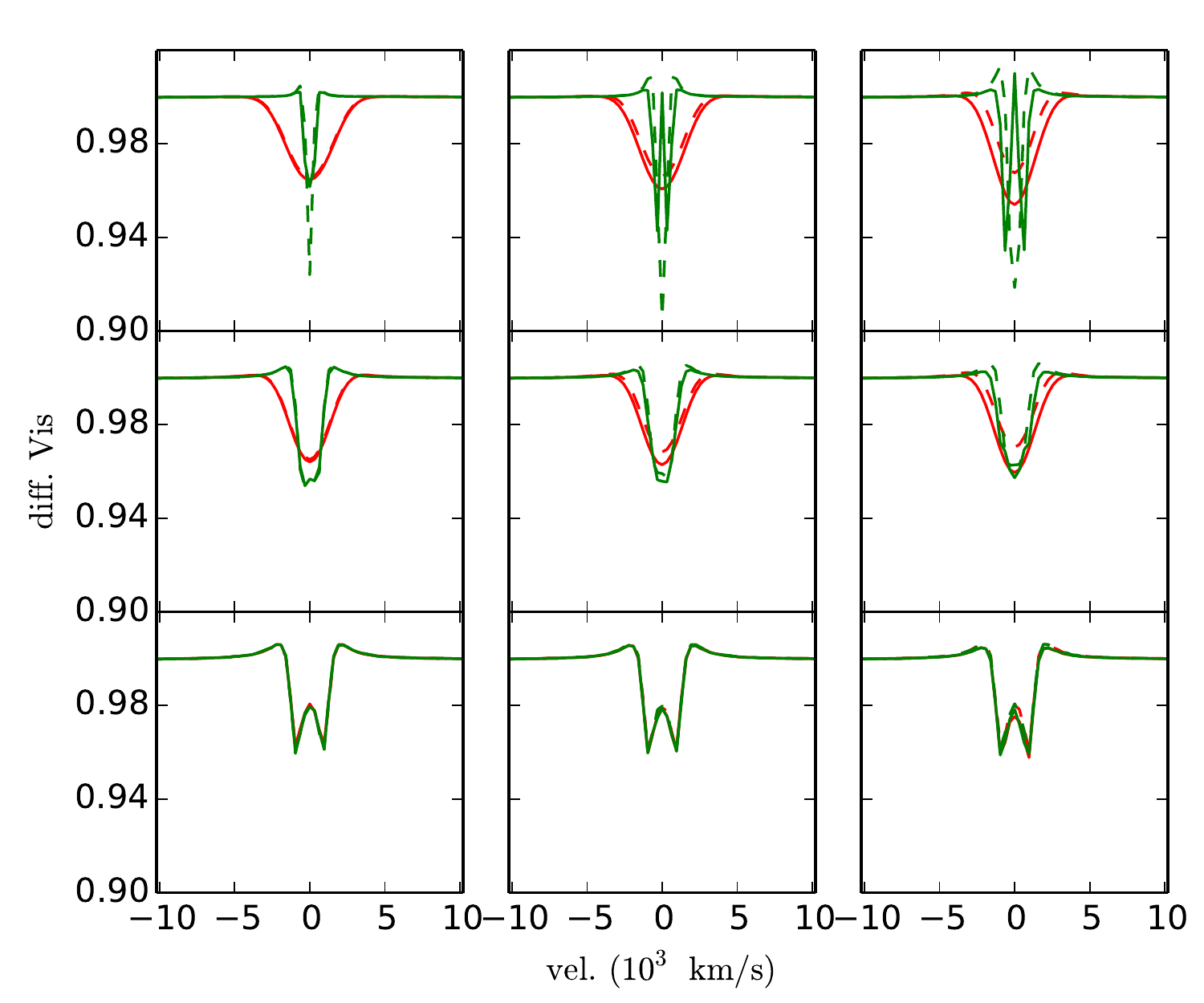}}  
  }  
 %\captionsetup{justification=justified, singlelinecheck=false} 
 \caption{Differential phase in degree (left panel) and differential visibility (right panel) in the $\parallel$ (dotted) and the $\perp$ (solid) baselines for the same grid as in Fig. \ref{Fig:gridSPECT-SITAU}.}\label{Fig:gridPHIVIS}
 \end{figure*}
%\newpage

%############################### Thick geometry ######################################
%############################# ---------------###########################
%############################### Thick geometry ######################################
%############################# ---------------###########################

\begin{figure*}
\centering
 \setlength{\unitlength}{1cm}
  \begin{picture}(18, 8)
  %\resizebox{9cm}{8cm}
  \put(0, 0){\includegraphics[width=18cm, height=8cm]{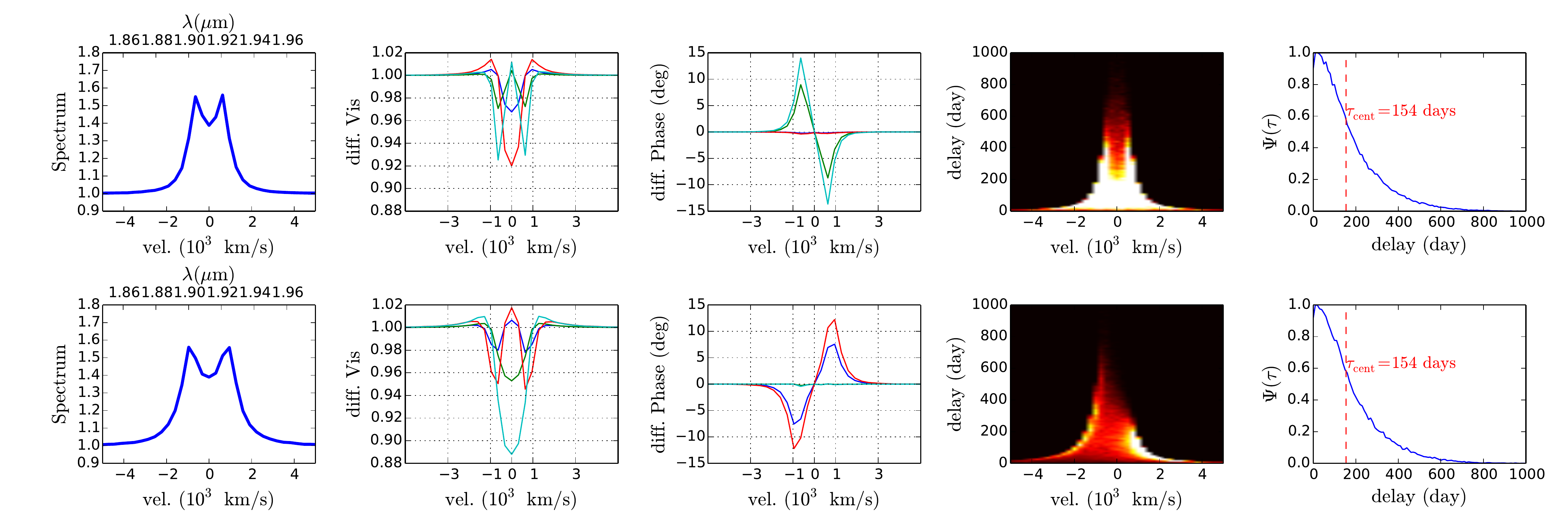}} 
  %\put(0, 4){\includegraphics[width=18cm, height=4cm]{\mypath/Kep_SMALL.pdf}}
  \put(1.3, 6.8){a. Keplerian}
  %\put(0, 0){\includegraphics[width=18cm, height=4cm]{\mypath/Inflow_SMALL.pdf}}
  \put(1.3, 3.0){b. Freefall}
  %\put(0, 0){\includegraphics[width=18cm, height=4cm]{\mypath/outflow_SMALL.pdf}}
  %\put(1.0, 3){c. Outflow}
    
  \end{picture}

\caption{Spectrum (left), differential visibility (left-middle), differential phase (middle), 2D response function (middle-right), 1D response function (right) is plotted for different velocity profiles and for flat geometry cases; Keplerian (upper row) and free fall inflow (lower row). OI observables are computed using U1 (red), U2 (cyan), U3 (blue) and U4 (green) baselines. %A free fall inflow and a Hubble type outflow with $V_0$ at $R_{in}$=10 km/s is considered for illustration.
}\label{Fig:thindisk}
\end{figure*}

\subsection{Fundamental geometrical and kinematics parameters}\label{sec:Thick}
       
%Various studies indicate $\mathrm{H}\beta$ emission line originates in a more spherical distribution i.e a thick disc whereas C iv comes from a flattened distribution i.e thin disc.
Now that we have estimated the interferometric angular size of the BLR we will constrain the three key parameters that describe the global BLR structure: $i$, $\omega$ and $\sigma_0$. \citet{2006A&A...456...75C}, \citet{2010MNRAS.409..591F} and \citet{2012MNRAS.426.3086G} have shown that these parameters dominate the RM scale factor $f$ and hence the virial BH mass estimate. This is illustrated by Fig. \ref{Fig:MassRatio} where the measured BH mass (left panel) and the scale factor $f$ (right panel) are plotted as a function of $i$ for various values of $\omega$. These values result from the velocity range $\Delta V$ estimated by the standard dispersion $\sigma_l$ of the line profile and the typical time lag $\tau_{\mathrm{cent}}$ obtained from our model with a fixed input mass and different values of $i$ and $\omega$. $f$ is the ratio of model input mass to the output mass $M_{\mathrm{out}}=c\,\tau_{\mathrm{cent}}{\Delta V}^2/G$. Fig. \ref{Fig:MassRatio}  shows that changes in $i$ and $\omega$ can introduce more than a factor 10 error on the mass estimate and shows how important it is to constrain these parameters.

%Later sections, with first step estimates of $i$, $\omega$ and  $\sigma_0$, we will consider other parameters, such as the nature of the global velocity field (rotation law, inflow, outflow), the contribution of macroturbulence to the local velocity field, and other geometric futures such as the anisotropic response of clouds related to their optical thickness.

%\citet{2012MNRAS.426.3086G} suggested “bowl shaped” geometry extended from outer edge of accretion disk to inner edge of dusty torus and BLR scale height changes by $R^2$ with existence of significant turbulent velocity.

%In our model, it is easy to transform the geometry from thin to thick by changing $\omega$.
Fig. \ref{Fig:gridSPECT-SITAU} shows the RM observables for a ($i-\omega$) grid with different line widths $\sigma_0$ considering a fixed BLR size $\sigma_{\mathrm{blr}}=0.4$ mas.  The spectra are plotted on the left panel and the 1D response functions are plotted on the right panel. The green curves show the spectra obtained with a fixed $\sigma_0=85$ km/s. The width of the spectrum is sensitive to the inclination and the opening angle. For small $\omega$, increasing $i$ clearly leads to the typical double peaked line profile of a Keplerian thin disk. Increasing $\omega$ broadens the line profile and blurs the double peaks until it forms flat top line profile, independent of $i$ as we approach a spherical structure with large $\omega$. The red curves represent the line profiles broadened by a change in $\sigma_0$ in order to obtain an equivalent global line width $\Delta V=3300$ km/s in all cases. The corresponding $\sigma_0$ is indicated in each picture. The $\sigma_0$ broadening blurs all the line details, but for the largest opening angles. The 1D delay transfer function $\Psi(\tau)$ is independent of $\sigma_0$. Its exact shape very slightly changes as a function of $i$ that shifts its maximum, and $\omega$ that makes the drop sharper for small delays. The RM BLR size $c\tau_{\mathrm{cent}}$ is not constrained by these parameters. The overall conclusion of this figure is that RM alone cannot separate $i$, $\omega$ and  $\sigma_0$ from $\Delta V$ and $\tau_{\mathrm{cent}}$ measurements only. However, a detail line profile analysis could discriminate these parameters up to a certain accuracy.
To show the effect of $i$, $\omega$ and  $\sigma_0$ on OI observables we plotted the differential phase (left panel) and the differential visibility (right panel) in Fig. \ref{Fig:gridPHIVIS} for the $\parallel$ (dotted) and the $\perp$ (solid) baselines. The photocenter shift between the line-emitting region and the continuum source increases with $i$, which increases the line of sight velocities. It globally decreases with $\omega$ that makes the iso-radial velocity regions more and more symmetric. Differential phase for a large opening angle shows sharp turns whereas the high local velocity case shape is much smoother and with reduced amplitude. An increase in $\sigma_0$, which blurs the iso-velocity zones, also decreases the differential phase amplitude, but for an identical amplitude, the $\phi_{\mathrm{diff}}(\lambda)$ function shows much sharper angles for a high $\omega$ than for a high $\sigma_0$.

Differential visibility is even sharper marker of the different models, if we have sufficient spectral resolution, i.e. sufficient SNR. %The global amplitude of the differential visibility drop can be explained by basically all combinations of parameters, as it is same in all pictures in Fig. \ref{Fig:gridPHIVIS}. 
In low spectral resolution, differential visibility is of little help. However, fine shape of the differential visibility spectacularly differs in different cases. Large opening angles yield a “w” shape that is independent from the direction of the baseline, while flat structures yield differential visibilities very sensitive to the baseline orientation, as it could be expected from Fig. \ref{Fig:line-Spect}, showing that the global size of the individual spectral bins is strongly different in the rotation axis and in the perpendicular direction. A large local velocity field removes this baseline direction dependence, but changes the curve shape and width.

%==========================================================================================================================

\subsection{Kinematics of the global velocity field}\label{sec:global}

Understanding the global kinematics of the BLR has been a long-standing problem as the sparsely sampled RM data was usually not sufficient to recover emission line as a function of velocity. However, recently various authors have found signatures of rotation, inflow or outflow in the BLR, analyzing high quality RM data and recovering $\Psi(v, \tau)$
\citep{2012ApJ...754...49P,2013ApJ...764...47G,2013arXiv1311.6475P}. On the other hand OI has been successful to provide signatures of rotation and expansion velocity in circum-stellar disks \citep{1996A&A...311..945S, 2007A&A...464...59M, 2012A&A...538A.110M}. To find constrains that OI can provide on the kinematics of BLR we simulated OI as well as RM observables.         
 
%In the previous section, we only considered the influence of the geometric parameters on the observable, here we will discuss the kinematics of the BLR.
%In the previous section we have discussed how it is possible to estimate the thickness of the geometrical distribution of clouds, its inclination and the contribution of the global velocity field. We have illustrated this with a global velocity field dominated by Keplerian rotation, but the discussion and the results would have been similar with a different global kinematics.
%Here we discuss the analysis of the kinematics in a system where we have already constrained the global geometry and the contribution of the global velocity field. As an example, we discuss this in the case of a flat BLR.					

Fig. \ref{Fig:thindisk}, shows the spectrum, the interferometric differential visibility and the differential phase together with the RM 2D and 1D response function for Keplerian rotation and free fall  kinematics models in a thin disk \citep[for details about the echo functions see][]{1991ApJ...379..586W}. We consider four baselines with different position angles:  U1 ($B=130\,\mathrm{m},\, PA=0\degree$), U2 ($B=130\,\mathrm{m},\, PA=90\degree$), U3 ($B=80\,\mathrm{m},\, PA=0\degree$) and U4 ($B=80\,\mathrm{m}, \, PA=90\degree$).

For a {\bf Keplerian rotation law}, as strongly suggested by Fig. \ref{Fig:line-Spect}, we see that for a baseline perpendicular to the rotation axis (baseline with $PA=90\degree$) the difference between the line and the continuum photocenter grows as we enter the line, cancels in the line center and reverses in the second half of the line. This gives a typical “S” shaped differential phase with an amplitude proportional to the resolution factor $\alpha$ defined by Eq. \ref{eq:vc}. In the direction of the rotation axis (baseline with $PA=0\degree$), the photocenter displacement and the differential phase are 0. The differential visibility globally displays a  “w” shape in the $\perp$ direction and a “v” shape in the $\parallel$ direction, with an amplitude depending on $\alpha$ and going from a peak over the continuum visibility (BLR smaller than the inner dust rim) to a visibility drop with a depth growing with $\alpha$. For an {\bf inflow}, the velocity amplitudes are larger for the same BH mass, as shown by the line profile and the 2D echo diagram, but this can be compensated by a mass change and hence introduces a mass uncertainty. The general shape of the curves is similar but the $\parallel$ and the $\perp$ directions are exchanged. The photocenter shift is now $\parallel$ to the rotation axis. The same exchange between the $\parallel$ and the $\perp$ directions can be observed on the differential visibility. The 2D echo diagram is different but this difference can be seen only on very high quality data. 

\begin{figure}
\centering
\includegraphics[width=8cm,height=6cm]{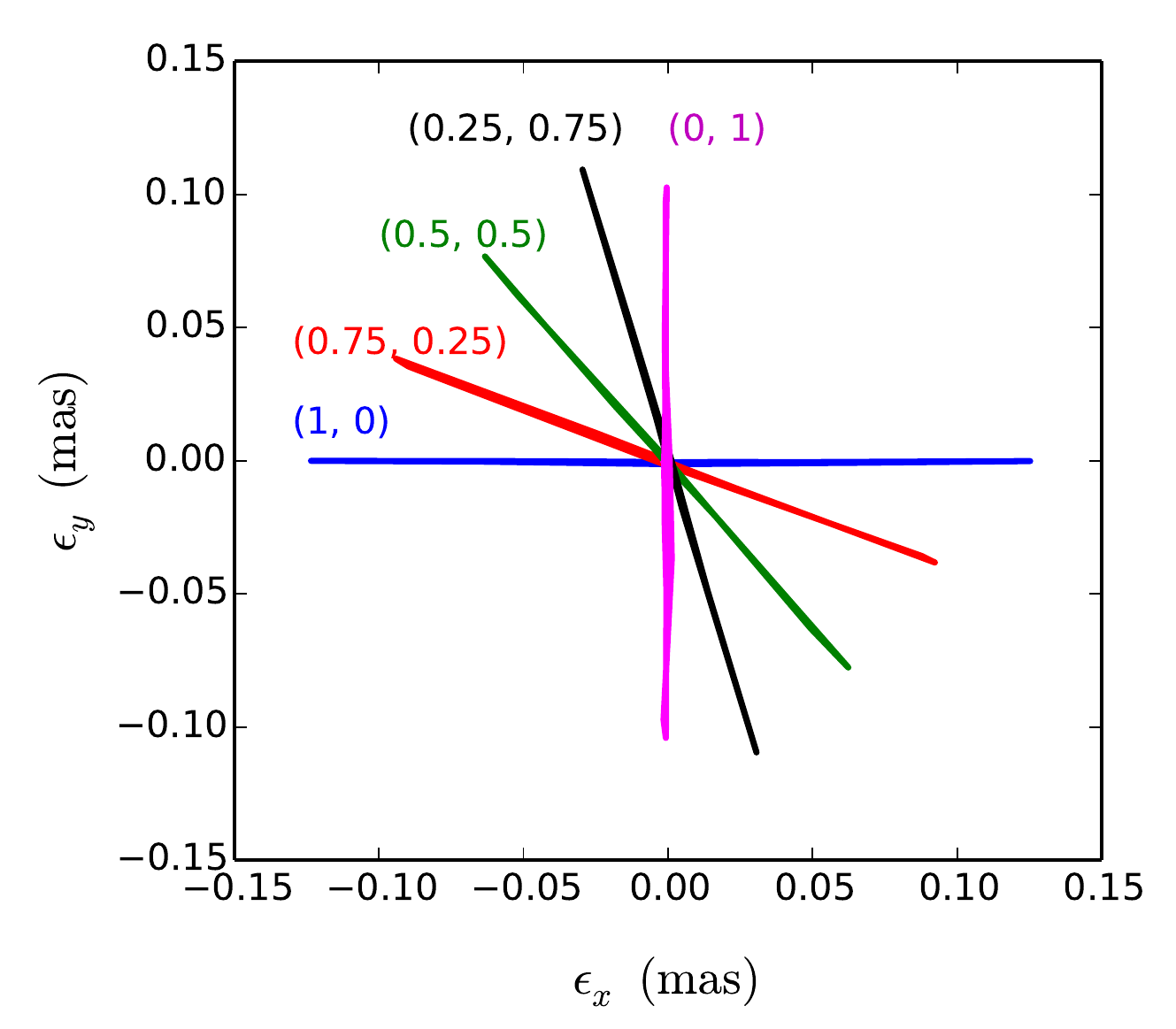} %or PhotoKI.pdf
\caption{Photocenter displacement in the sky plane for different combinations ($V_k$, $V_f$) of Keplerian ($V_k \times V_a$) and Freefall velocity ($V_f \times V_c$) amplitude. The photocenter for pure Keplerian case is represented in blue whereas pure Freefall is presented in magenta. In figure, y is the projected direction of the symmetry axis i.e. position angle $\Theta=90$.}\label{Fig:PHOTOKI}
\end{figure}

The decisive capacity of differential measurements with OI to discriminate between rotation and inflow/outflow is further illustrated by Fig. \ref{Fig:PHOTOKI} that shows the rotation of the global photocenter shift with $\lambda$ as the ratio between rotation and expansion changes, as first illustrated by \citet{1992ESOC...39..403C} for circumstellar disks. Here the y direction is defined by the projected axis of symmetry i.e. the position angle of the disk $\Theta=90\degree$. In this context, \citet{1996A&A...311..945S} has shown that the trajectory of the photocenter displacement vector $\vec{\epsilon}(\lambda)$ yields the strongest constraint on the velocity law index $\beta$ and \citet{2007A&A...464...59M,2012A&A...538A.110M} has shown that the equatorial disk of Be stars is strongly dominated by a Keplerian rotation ($\beta=0.5$). The same approach can be applied to disk BLRs.

%The differential visibility shows a rise (Fig. 3) on the top continuum for a compact BLR that resides inside the dust continuum region and the size can also be obtain from the response function centroid. The differential phase amplitude is decreased accordingly keep the shape constant as obtained for large extended BLR case. 

%=================================================

%######################## Macroturbulence ##############################
%######################## ---------------###############################

\subsection{Macroturbulence}\label{sec:macro}
%Large relative flux of ionizing continuum and emission line suggest that BLR clouds should have significant scale height and for a thin disk the ratio $H/R \simeq 0.1$, where $H$ is the scale height at radius $R$.
%The standard accretion disk model by \citet{1973A&A....24..337S} suggest to have flared disk where scale height increases more rapidly with radius $R$. 

%In sec \ref{sec:Thick} we have explained that we have to separate the global velocity field from the local one. This local velocity field was represented by the rest line width $\sigma_0$, whose broadening should result from gas motion inside the clouds. 

Several authors have introduced models where the dynamics of the clouds is dominated by random macroturbulence \citep{2006A&A...456...75C,2008MNRAS.390.1413F,2010MNRAS.409..591F,2012MNRAS.426.3086G,2014arXiv1407.2941P}. Macroturbulence in the BLR can provide the internal pressure required to support the disk vertical extent \citep{1973A&A....24..337S}. %Recent dynamical model of BLR in Mrk 50 of \citet{2012ApJ...754...49P} suggest a flared disk geometry for this object with an opening angle $25\pm 10\degree$ that corresponds to a scale height $H\sim 0.5$ at the outer radius. 
\citet{2006A&A...456...75C} suggested various disk geometries and implemented a turbulent velocity that depends on the scale height of disk. \citet{2012MNRAS.426.3086G} showed that for an object of low inclination, macroturbulence dominates the Keplerian velocity and can produce significant broadening. 
%The clouds at large radius responds more and hence shows Lorentzian line profile. 

\begin{comment}
Its direction is random and its amplitude depends on the radius and on the local thickness of the BLR that in our model is represented by the opening angle $\omega$:
\begin{equation}
|V_{\mathrm{turb}}|=V_{rot}\,P_{\mathrm{turb}}\,r\,\mathrm{sin}\,\omega
\end{equation}
The multiplicative factor $P_{\mathrm{turb}}$ is a parameter that must be fitted from the data to constrain macroturbulent physical models. The macroturbulence is zero both for flat disks ($\omega=0$) and for $P_{\mathrm{turb}}=0$. 
\end{comment}

%However a thermal microturbulence velocity can be present inside the cloud and it mainly defines the shape of local line width($\sigma_{line}$). In our model because of large number of clouds these two effects are equivalent and we choose a fixed $\sigma_{line}$. 

\begin{figure}
\centering
\resizebox{8.6cm}{7cm}{\includegraphics{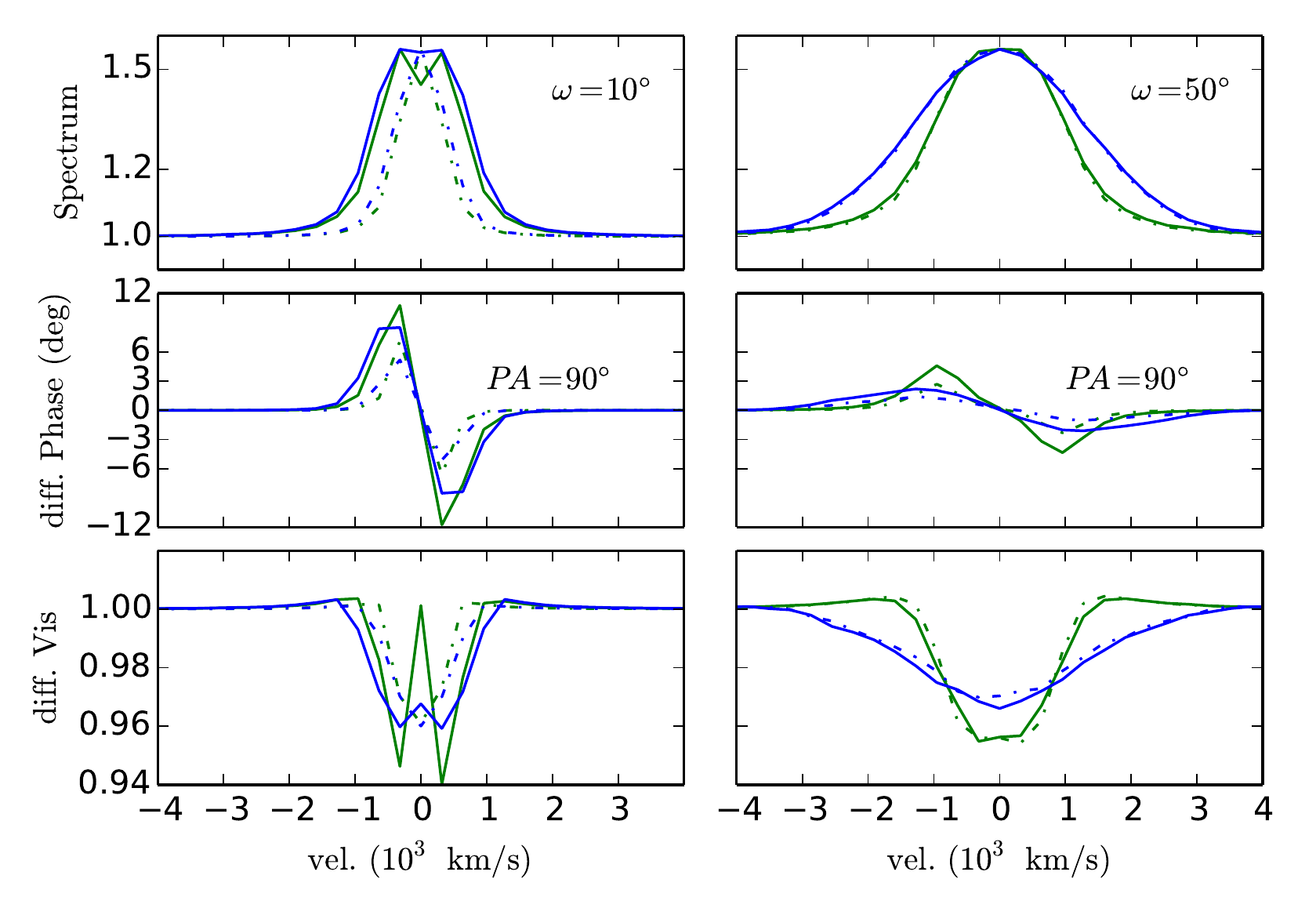}}
\caption{The effect of macroturbulence on different observables for two different inclinations $10\degree$ (dash-dot line) and $20\degree$ (solid line), and two different opening angles $10\degree$ (left column) and $50\degree$ (right column). The macroturbulence parameters $P_{\mathrm{turb}}=0$ is plotted in green whereas $P_{\mathrm{turb}}=4$ is plotted in blue.}\label{Fig:turb}
\end{figure}

We use a similar approach and introduce a macroturbulent velocity component in our model defined by Eq. \ref{eq:macro}. Fig. \ref{Fig:turb} shows the effect of macroturbulence on the spectrum (upper), the differential phase (middle) and the differential visibility (lower) in the $\perp$ baseline, for different opening angles $\omega$ and turbulence parameters $P_{\mathrm{turb}}$. Macroturbulence broadens the line profile and enhances the response of the line wings. From the general shape of all the other observables, it is impossible to discriminate between the effect of an increase of the local line width $\sigma_0$ and that of an increased macroturbulence. However, even if we cannot tell if the local velocity field is dominated by $\sigma_0$ or by macroturbulence, we can separate it from the global velocity field and therefore obtain all the global geometric and kinematic parameters. 

Neither RM nor DI can allow an experimental separation of the effects of the local line width and of the macroturbulence as long as we analyze them in terms of equivalent broadening of $s(\lambda)$, $V(\lambda)$, $\Phi(\lambda)$. In all cases it should affect the balance between the line wings and the line core (the ratio between FWHM and standard dispersion $\sigma_l$ of the line profile) in a similar way for all observables. A more advanced physical analysis of the local line profiles and of the macroturbulent velocity statistics could yield parameters that could be discriminated  by advanced model fits of RM and DI data with very high SNR. However, for the present study, the key point is that they both can be represented by a single parameter (either $\sigma_0$ or  $P_{\mathrm{turb}}$) whose choice does not have any impact on the other parameter estimates.

%{\bf In the spectral line profile, macroturbulence mainly changes the weight of the line wings. The ratio between the line FWHM and its standard dispersion $\sigma_l$ increases when $\omega$ increases or $i$ decreases (upper panel of Fig. \ref{Fig:turb}). For known $i$ and $\omega$, this ratio $\mathrm{FWHM}/\sigma_l$ could be used to constrain the relative contributions to the local velocity field. However, this ratio is also sensitive to geometrical parameters such as the radial distribution of clouds as well as to their exact spectral response. At a large radius the clouds have a stronger response with a Lorentzian line profile according to \citet{2012MNRAS.426.3086G}. A fine analysis of the effect of macroturbulence needs a physical model of the cloud response that is beyond the scope of this paper.}

%{\it The observables do not allow discrimination between macroturbulence and other contributions to the local velocity field and this discrimination is not necessary to interpret the other parameters.}

%######################### Effect of Anisotropy #############################
%######################### --------------------#############################

\subsection{Anisotropy}\label{sec:anis}
\begin{comment}
\begin{figure}[h]
\centering
\includegraphics[width=8cm,height=6cm]{\mypath/PHOTO.pdf}
\caption{Shift of photocenter in the sky plane for different $F_{\mathrm{anis}}$ in the case flat Keplerian geometry.}\label{Fig:photocenter}
\end{figure}
\end{comment}

The emission line optical depth of each cloud determines the anisotropy of the re-emitted light, from an isotropic emission for an optically thin cloud to a maximum anisotropy, with dark and bright sides for a thick cloud. \citet{1994MNRAS.268..845O} computed this for different strong emission lines and suggested that lines with a large ionization parameter are emitted anisotropically at some radii. 
%\citet{2012MNRAS.426.3086G} has calculated the effect of anisotropy in the their bowl shape geometry and found a stronger dependency of inclination compare to the thin disk geometry. 
Anisotropy increases the time lag almost without changing the line profile, which can be a cause of error for the mass estimate \citep{1996ApJ...469..113G}. 

\begin{figure}
\centering
\includegraphics[width=8.6cm,height=4.0cm]{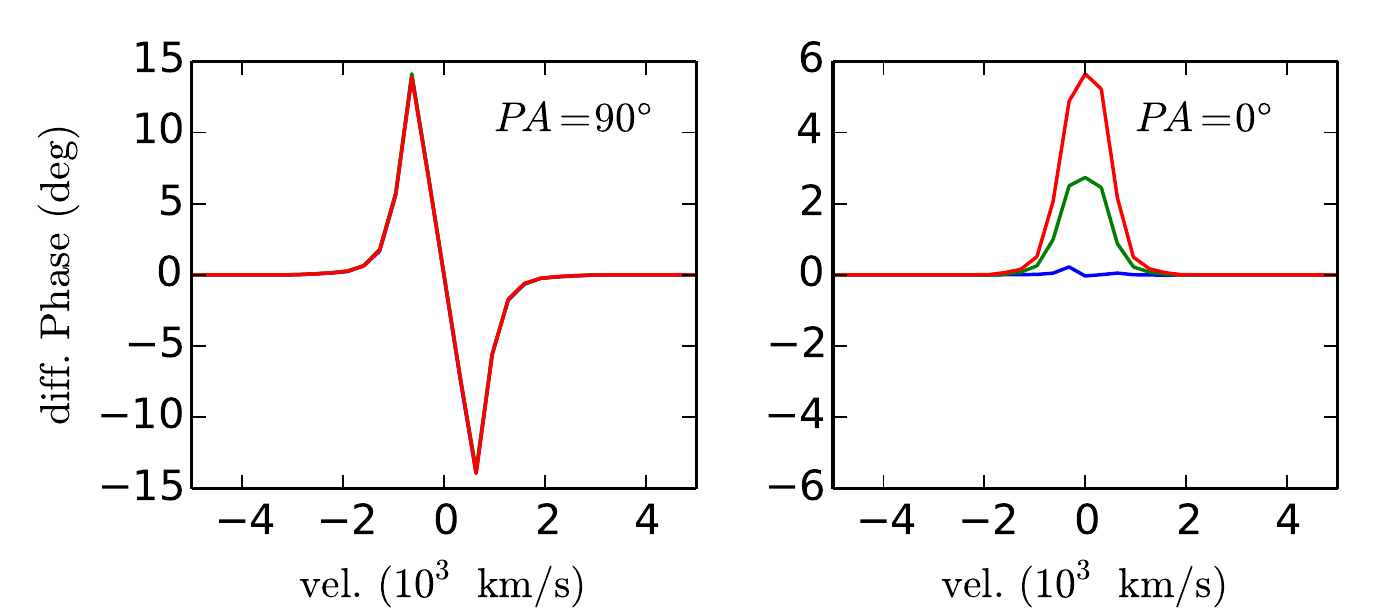}
\caption{Differential phase in the $\perp$ (left panel) and the $\parallel$ (right panel) baselines for a flat Keplerian disk is plotted for various anisotropic cases: $F_{\mathrm{anis}}$=0 (blue), 0.5 (green) and 1.0 (red).}\label{Fig:ani}
\end{figure}

To compute the effect of anisotropy, we use Eq. \ref{eq:anis} and its effect on the differential phase is shown in Fig. \ref{Fig:ani}. We see a strong effect in the direction of the rotation axis and no effect in the $\perp$ direction. This is because the inclusion of anisotropy reduces the emission on the near-side and increases to the far-side, enhances its brightness and shifts the photocenter towards its direction. The photocenter along the $\parallel$ direction changes rapidly while the photocenter in the $\perp$ direction remains unchanged. Differential phase is therefore a good marker of anisotropy, particularly if we have a priori information on the axis position angle.

%\footnotemark[17]

%========================  TABLE for fringe detection and differential observation plot==============
\begin{center}
 \begin{table*}
 %\resizebox{18cm}{!} {
	\caption{Parameters for fringe detection limit$^a$ and differential observation of BLR$^b$ plot.}
	\begin{minipage}{\textwidth}  
 	\centering {
 	%\small\addtolength{\tabcolsep}{4pt}
     \begin{tabular}{ |c |c| c| c| c| c| c| c|c| c| c| c| c| c|}  \hline
    
     \multicolumn{1}{ |c| }{ \bf{Instrument}} 
      & \multicolumn{13}{ c| }{ \bf{Parameters}} \\ \cline{2-14}
     %Instrument & Parameters \\ \hline
     \multicolumn{1}{|c|}{} & $n_T$ & $t_{\mathrm{DIT}}$ (sec) & $n_p$ & $\sigma_{\mathrm{RON}}$ & $n_{\mathrm{th}}$ & $V_{\mathrm{inst}}$ & $n_\lambda$ & $N_{\mathrm{EXP}}$ & $n_0$ & A ($\mathrm{cm}^2$) & $S$ & $T$ & $R$
     \footnote[3]{Resolution per spectral pixel. The actual spectroscopic resolution is typically $R/2$.}
      \\ \hline
     AMBER    & 3		& 0.2	& 32 & 	11	& 1.07	& 0.25 	&  512	& 100$^a$, 36000$^b$ & 45	& 497628 & 0.5	& 0.01		& 3000   \\ \hline
     AMBER+   & 3		& 0.2	& 32 & 	11	& 1.07	& 0.25	&  512	& 100$^a$, 36000$^b$ & 45	& 497628 & 0.5	& 0.01		& 3000   \\ \hline
     OASIS    & 3		& 0.2	& 4	 &  11	& 1.07	& 0.25	&  512	& 100$^a$, 36000$^b$ & 45	& 497628 & 0.5	& 0.01 $\times$ 7	& 3000  \\ \hline
     OASIS+   & 3		& 0.1	& 4	 & 	3	& 1.07	& 0.5   &  512	& 200$^a$, 72000$^b$ & 45	& 497628 & 0.5	& 0.01 $\times$ 7	& 1000   \\ \hline
     GRAVITY  & 4	    & 60  & 4
     \footnote[4]{GRAVITY has a pair wise beam combiner and must be analyzed as a 2T interferometer, with 1/3 of the flux in each aperture.} & 11 	& 1.07	& 0.75	&  512	& 120$^b$			 & 45	& 497628 & 0.5	& 0.01	& 1000   \\ \hline
     OASIS+FT   & 3		& 60 	& 4	 & 	3	& 1.07	& 0.75   &  512	& 120$^b$	& 45  & 497628 & 0.5	& 0.01 $\times$ 7	& 1000   \\ \hline
 	\end{tabular} \label{TABLE:2}
 }
 \end{minipage}
 \end{table*}
 \end{center}
%###################Interferometric observation of BLR #########################
%###################-----------------------------------#########################

\section{Feasibility of interferometric observation of BLR}\label{sec:present and future}
The potential of OI depends on the number of objects that can be observed. As the VLTI is the only interferometer having near-infrared instruments with a high enough spectral resolution (i.e. $R \geq 500$) to spectrally resolve BLR emission lines because of its 8-m Unit telescopes (UTs), we evaluate this number for current, upcoming and possible VLTI instruments. Other interferometric facilities such as CHARA or NPOI are equipped with substantially smaller telescopes.

%The successful observation of 3C273 influence us to do feasibility study of BLR observation with the current, upcoming and possible spectro-interferometric instruments in the near-infrared (near-IR) in order to see the number of object that could be accessible with VLTI to set the possibility of having a large unification scheme. Actually, the VLTI is the only interferometer having near-infrared instruments with a high enough spectral resolution (i.e. $R \geq 500$) to spectrally resolve BLR emission lines. Moreover, we believe that the 8 m Unit telescopes (UTs) are a key feature for an AGN program with such spectral resolution. Other interferometric facilities such as CHARA or NPOI are equipped with substantially smaller telescopes.

In the case of each VLTI instrument, we examine the possibility to observe a target, i.e. to detect and maintain the fringes on the target itself; and the accuracy of its absolute visibility, differential visibility and differential phase.

\subsection{Current, incoming, and possible VLTI instruments }
Our feasibility analysis has been made for the following instruments:
%\begin{itemize}

%\item
{\bf AMBER} \citep{2007A&A...464....1P} is the first-generation near-IR spectro-interferometric VLTI instrument. With its standard frame-by-frame processing, it cannot observe AGN in medium resolution (MR), as the current VLTI fringe tracker, used to stabilize fringes, is limited to about $K<9$. However, AMBER can already operate in an alternative mode, called AMBER+ \citep{2012SPIE.8445E..0WP}, where the full dispersed fringe image is processed, in a way equivalent to a coherent integration of all spectral channels, whatever the SNR per channel is. On the QSO 3C273 in MR ($R=1500$), the fringes were detected with a SNR=3 in typically 1 s. To obtain differential visibility and phase with a sufficient accuracy (respectively 0.02 to 0.03 and $1\degree$ to $2\degree$) it has been necessary to bin the spectral channels down to a resolution 250 as we had only $\sim 1$ hour of actual observation. The results achieved with AMBER+ on 3C273 have been used to validate our SNR computations.

%\item 
{\bf OASIS} (“Optimizing AMBER for Spectro-Interferometry and Sensitivity”) is a light AMBER modification proposed by (Petrov, 2014) that could be installed in a few months if ESO opens few days slot in the VLTI planning. It uses spectral encoding to separate the fringe peaks, allowing to code the interferogram on 4 pixels instead of the 32 pixels currently used in AMBER. Moreover, the spatial filters with fibers, which are not critical for differential observables, would be bypassed by optimized optics that yields a gain in transmission by a factor $\sim$ 7 with respect to the current AMBER instrument. OASIS would use the current spectrograph and detector of AMBER.

%\item
{\bf OASIS+} would be a major improvement of AMBER+ with OASIS. It would use a specific spectrograph optimized for BLR observations with a fixed resolution of 1000 and this would allow using a SELEX detector with much lower read out noise. It could be developed as a visitor instrument in the 1 M\euro{} range. We note that OASIS+, or any other successor of the 2nd generation VLTI instruments, is not in the current ESO plans, but it gives an idea of what could be ultimate VLTI performance for AGNs. 

%\item 
{\bf GRAVITY} \citep{2008SPIE.7013E..69E} is the second-generation VLTI near-IR spectro-interferometric instrument that will be commissioned around 2016. Its main characteristic of interest for a BLR program is that it has an internal fringe tracker that should allow cophased observations up to $K=10.5$. This allows much longer individual frame time, a higher instrumental visibility and a more stable one. The current GRAVITY plans do no foresee using it without its fringe tracker, but an AMBER+ type of processing is possible and would allow performances intermediate between AMBER+ and OASIS for $K>10.5$.

%\item 
{\bf OASIS+FT} refers to the use of OASIS+ with a second-generation fringe tracker (FT), with a limiting sensitivity larger than $K=10.5$. Such a FT would allow increasing the accuracy of the measurements just like the one in GRAVITY, and it would also extend the possibilities of GRAVITY. Currently proposed designs show that FT magnitudes higher than 13 in $K$-band should be achievable \citep{2012SPIE.8445E..1LM, 2014SPIE.9146E..2P}. The development of such a FT should be a priority for an extended AGN program with the VLTI.

%\end{itemize}

Table \ref{TABLE:2} summarizes the observing parameters for these different instruments and modes.

%\begin{comment}

\subsection{Interferometric signal and noise}\label{sec:ref}

From a general formalism described in \citet{1986JOSAA...3..634P} and updated in \citet{2006MNRAS.367..825V}, it is easy to show that the noise on the coherent flux computed from each all-in-one multi-axial interferogram (like in AMBER) is given by:
\begin{equation}
\sigma_c=\sqrt{n_Tn_*t_{\mathrm{DIT}} + n_p\sigma_{\mathrm{RON}}^2 + n_T n_{\mathrm{th}}t_{\mathrm{DIT}}},
\end{equation}
 where  $n_*$ is the source flux per spectral channel, frame and second,  $n_T$ is the number of telescopes, $t_{\mathrm{DIT}}$ the frame exposure time, $n_p$ is the number of pixels (for multi-axial instruments up to $n_T=4$ we have $n_p=4n_{T}(n_T-1)/2$),  $\sigma_{\mathrm{RON}}^2$ is the variance of the detector read-out noise and $n_{\mathrm{th}}$ is the background flux per spectral channel, frame and second. In $K$-band this value is much smaller than the detector noise and hence can be neglected for short exposures. However for long exposures such as in cophased mode $n_{\mathrm{th}}$ should be taken into account. In $K$-band, $n_{\mathrm{th}}=1.07$ photons $\mathrm{s}^{-1} \mathrm{cm}^{-2} \mu \mathrm{m}^{-1}$. For a pair-wise instrument like GRAVITY, the same formula applies with $n_p=4$ and the flux of each telescope has to be divided by $(n_T-1)$ that is the number of pairs each aperture is involved in.

The classical SNR on the coherent flux, per spectral channel and per frame \citep{2006MNRAS.367..825V, 2012SPIE.8445E..2JL} is then given by:
\begin{equation}
\mathrm{SNR}_{0}=\frac{C}{\sigma_c} \simeq \frac{n_*t_{\mathrm{DIT}}V_{\mathrm{inst}}}{\sigma_c},
\end{equation}
where $V_{\mathrm{inst}}$ is the instrument visibility. 

The source flux per spectral channel per frame and per second is given by:
 \begin{equation}
 n_{\star}=n_0AST\delta{\lambda}10^{-0.4K_{\mathrm{mag}}},
 \end{equation}\\
 where $n_0$ is the number of photons per second per $\mathrm{cm}^{2}$ per $\mu \mathrm{m}$ from a star with $K_{\mathrm{mag}}=0$, outside earth atmosphere, $n_0=45 \times 10^6$ photons $\mathrm{s}^{-1} \mathrm{cm}^{-2} \mu \mathrm{m}^{-1}$, $A$ is the collecting area of telescope, $S$ is the Strehl ratio with the VLTI adaptive optics system MACAO, $T$ is the overall transmission of the atmosphere, the VLTI and the instrument, and $\delta \lambda$ is the spectral band width $=\lambda_0/R$, where $R$ is the resolution.
 \\
 \\
{\bf Standard processing}: As the coherent flux C is affected by a random atmospheric phase, we have to average its squared modulus $|C|^2$ over all available spectral channels and several frames. The SNR of such a quadratic average is given by:
 \begin{equation}
 \mathrm{SNR_{std}}=\frac{\mathrm{SNR}_0^2}{\sqrt{1+2\mathrm{SNR}_0^2}}\sqrt{N_{\mathrm{EXP}}n_\lambda},
 \label{eq:snr_std}
 \end{equation}
 where $n_\lambda$ is the number of spectral channels and $N_{\mathrm{EXP}}=\frac{t_{\mathrm{EXP}}}{t_{\mathrm{DIT}}}$ is the total number of $t_{\mathrm{DIT}}$ frames processed in the $t_{\mathrm{EXP}}$ total time.
\\
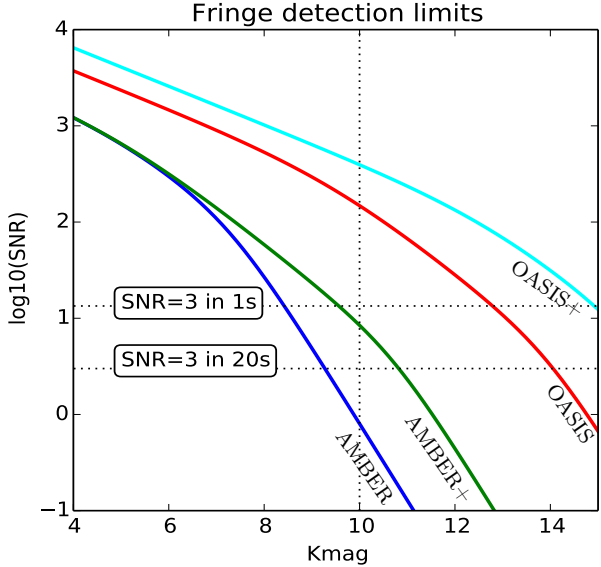
\begin{figure}
 \centering
 \setlength{\unitlength}{1cm}
 \begin{picture}(9, 8)
 %\resizebox{9cm}{8cm}
 \put(0, 0){\includegraphics[width=9cm, height=8cm]{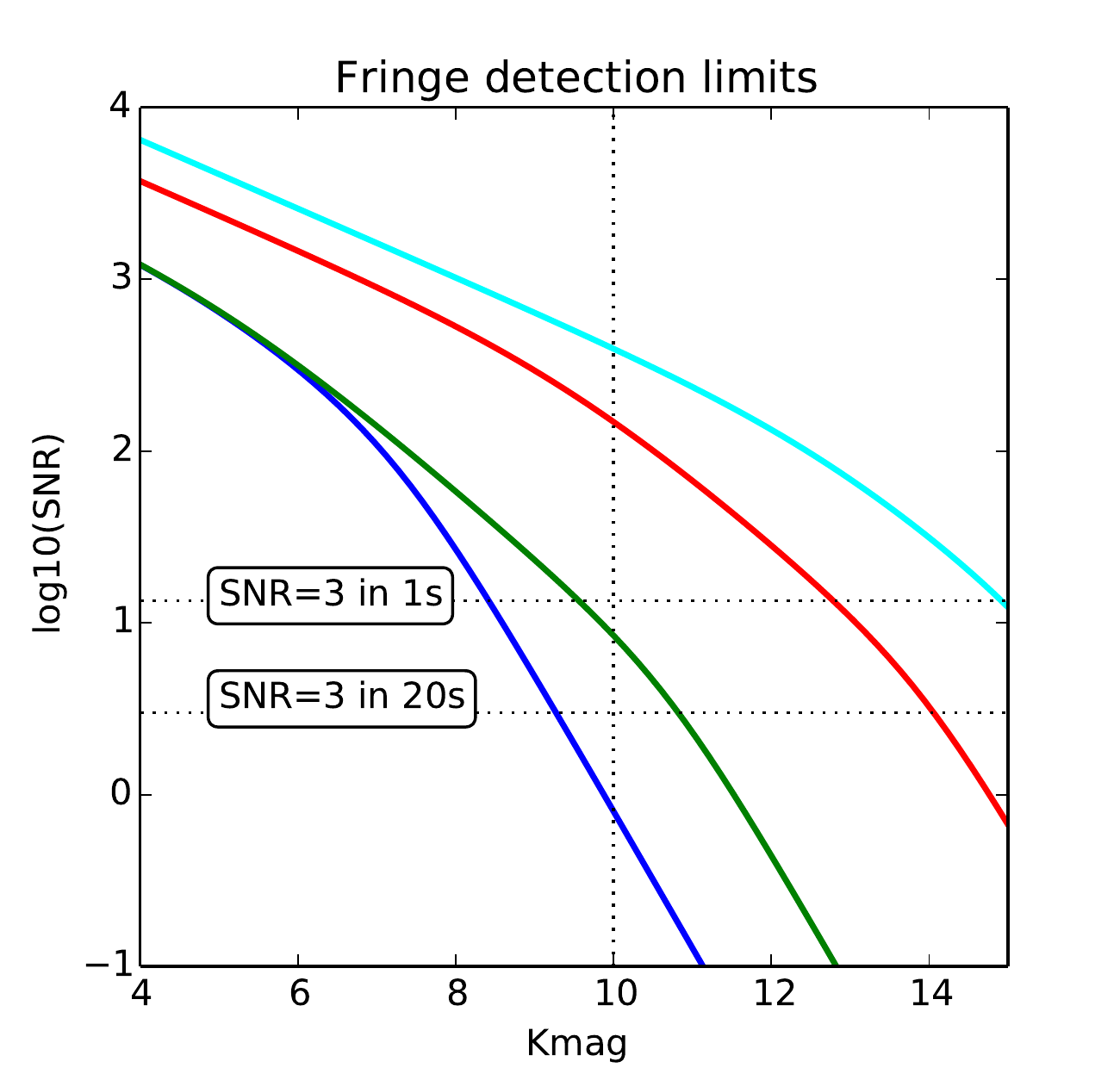}}
 \put(4.6, 1.8){\rotatebox{-55}{AMBER}}
 \put(5.5, 2.0){\rotatebox{-55}{AMBER+}}
 \put(7.4, 2.4){\rotatebox{-55}{OASIS}}
 \put(6.9, 4){\rotatebox{-35}{OASIS+}}
 \end{picture}
 \caption{Fringe detection (log10(SNR)) limits for different VLTI instruments. From left to right: standard AMBER performance with 0.2 s frames (blue), current AMBER+ measured performance with incoherent TF2D processing (green), OASIS module (red) and OASIS+ module (cyan). The AMBER+ curve (given here for a maximum of 20 s) is compatible with our experimental result of fringe detection with SNR=3 in 1 s. The horizontal dotted black line shows the threshold fringe detection limit of SNR=3 in 1 s and 20 s. The vertical dotted black line corresponds to $K_{\mathrm{mag}}=10$.}\label{Fig:SNR}
\end{figure}
\\
{\bf AMBER+ processing}: We have developed a new approach where the full-dispersed fringe image is processed, in a way equivalent to a coherent integration of all spectral channels, whatever the $\mathrm{SNR}_0$ per channel is. This data processing is explained in \citet{2012SPIE.8445E..0WP}. Then we still have to make a quadratic average of the other frames and the SNR of this processing is given by
 \begin{equation}
  \mathrm{SNR}_{+}=n_\lambda\frac{\mathrm{SNR}_0^2}{\sqrt{1+2{n_\lambda}\mathrm{SNR}_0^2}}\sqrt{N_{\mathrm{EXP}}}.
  \label{eq:snr+}
 \end{equation}
 This allowed a gain of typically $\sqrt{n_\lambda}$ which made possible the first observation of 3C273 with a spectral resolution $R=1500$. The fringes were detected with a typical $\mathrm{SNR_{+}}=3$ in 1 s exposures.

The phase is estimated from the average coherent flux and its accuracy is given by
 \begin{equation}
 \sigma_{\phi}=\frac{<C>}{\sigma_c \sqrt{2\, n_b}}=\frac{1}{\mathrm{SNR}_0 \sqrt{2\, n_b}\sqrt{N_{\mathrm{EXP}}}}
 \end{equation}
 with $N_{\mathrm{EXP}}=36000$ for 2 hours of observations and $n_b$ is the number of binning.
 
In AMBER+, a SNR analysis \citep{2014SPIE.9146E..2P} shows that 
 \begin{equation}
 \sigma_{\phi+}=\sigma_{\phi}\sqrt{2\frac{\sigma_{\phi}^2}{n_\lambda} + \frac{1+n_\lambda}{n_\lambda}}. 
 \label{eq:phiAMBER+}
\end{equation}

    \begin{figure*}
      \centering
       \setlength{\unitlength}{1cm}
        \begin{picture}(18, 8)
        %\put(0, 0){\includegraphics[width=9cm, height=8cm]{\mypath/target_phi3.pdf}}  % or  target_phi2.pdf was in submitted version 
 	    %\put(9, 0){\includegraphics[width=9cm, height=8cm]{\mypath/target_vis2.pdf}}  % or 
 		\put(0, 0){\includegraphics[width=18cm, height=8cm]{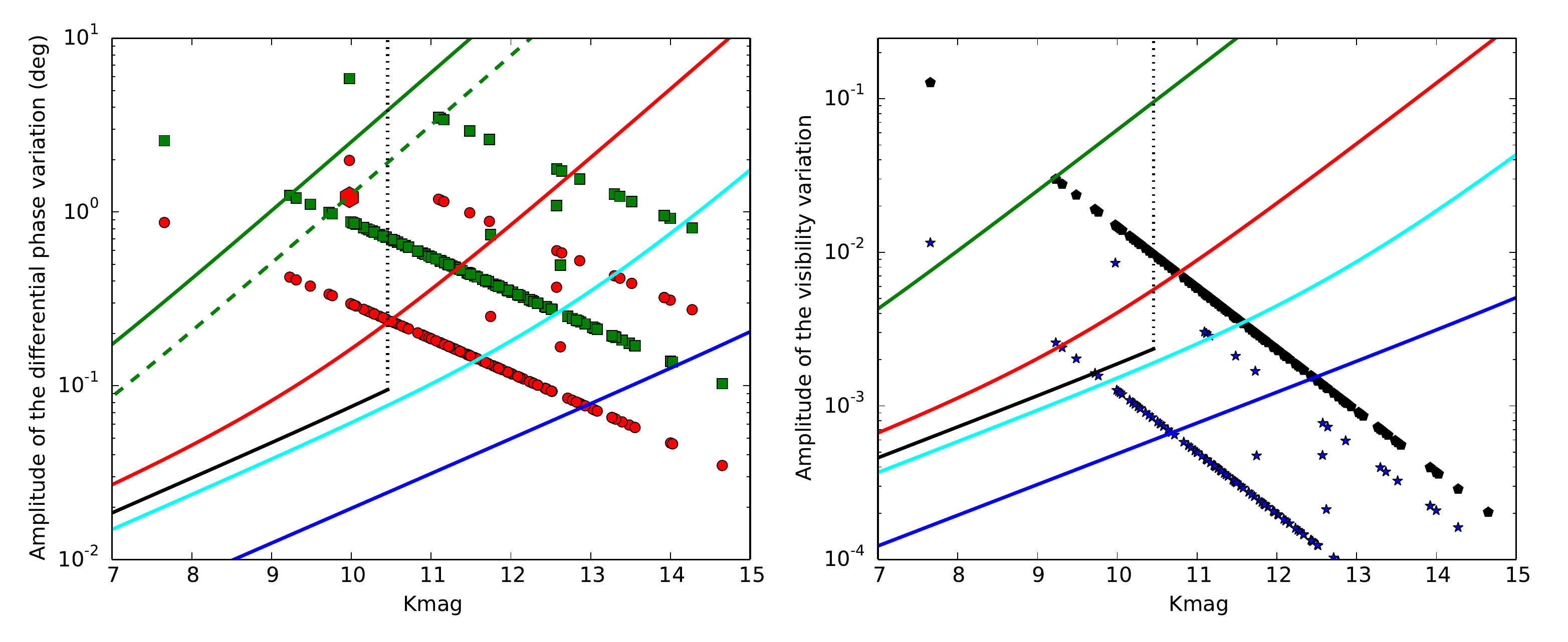}} 	
        \put(2.5, 6.1){\rotatebox{45}{3C273}}
        \put(3.1, 6.4){\vector(1, -1){0.8}} 
        \put(3.6, 6.3){\rotatebox{45}{AMBER+}}
        \put(7.0, 6.7){\rotatebox{45}{OASIS}}
        \put(7.6, 5.3){\rotatebox{45}{OASIS+}}
        \put(2.0, 1.2){\rotatebox{25}{GRAVITY}}
        \put(2.4, 1.6){\vector(-1, 1){0.4}}
        \put(4, 1.2){\rotatebox{25}{OASIS+FT}} 
        \end{picture}
      
      \caption{Feasibility of observation of BLR with maximum VLTI 135 m baseline. The  left image shows the accuracy of the differential phase in degrees, for the different instruments described in the text (full lines labeled by the instrument name). The differential phase error is dominated by fundamental noise at least down to $0.05\degree$. The points represent the expected differential phase for all Sy1 AGNs and QSOs brighter than $K=15$ observable in Paranal ($\delta<15\degree$). Each source is represented by a red circle, corresponding to the minimum differential phase signal from the RM BLR size estimates, and a green square, corresponding to the maximum differential phase for a skewed continuum image. The 3C273 point (red octagon) represents the actual differential phase accuracy obtained with AMBER+. This value is on the dotted green line corresponding to the actual spectral channel binning and observation time on 3C273. The right image represents the differential visibility accuracy from the fundamental noises for different instruments (full lines). The systematic error on differential visibility is below $10^{-3}$. For instruments used without a fringe tracker (AMBER+, OASIS, OASIS+), the absolute visibility accuracy cannot be better than 0.03. For instruments operated with a fringe tracker (GRAVITY, OASIS+FT) we expect to go below 0.003. The points represent the absolute visibility (black polygons) and the differential visibility (blue stars) expected for each source, according to the inner rim size estimated from its luminosity. The horizontal scale gives the estimated $K$ magnitude of the central source.}
      \label{Fig:target}
    \end{figure*}
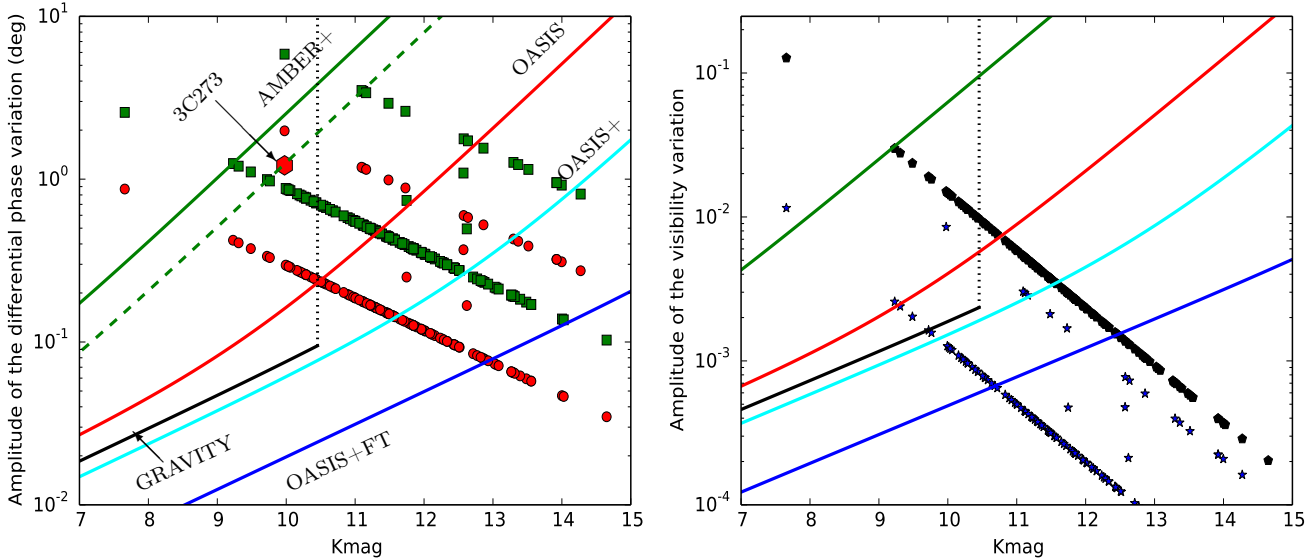
    
 %###################### Fringe detection limit ###################################
 \subsection{Fringe detection limit}\label{sec:SNR}
 
In Fig. \ref{Fig:SNR} we plotted the fringe detection limit $\mathrm{log_{10}\,(SNR)}$ as a function of $K_{\mathrm{mag}}$ using Eq. \ref{eq:snr_std} and Eq. \ref{eq:snr+} for  different instruments like standard AMBER performance with 0.2 sec per frame, AMBER+ performance with incoherent TF2D processing, OASIS module and OASIS+ module. The parameters used for the calculation are presented in Table \ref{TABLE:2}. We found AMBER+ could reach up to $K_{\mathrm{mag}}\sim 10.5$ and the potential limit of the new OASIS and OASIS+ $>13$.

We note that fringe detection limit for GRAVITY was not included in that computation as GRAVITY will use an internal fringe tracker working up to $K_{\mathrm{mag}}$=10.5. Nevertheless, it should be possible to use GRAVITY without the internal fringe tracker in a mode similar to AMBER+ mode, allowing to observe beyond $K_{\mathrm{mag}}$=10.5, with performances intermediate between AMBER+ and OASIS.

 %###################### Targets and amplitude of signals #########################

 \subsection{Signal estimation} \label{sec:signal_estimation}

In the following, we estimate three interferometric quantities: the absolute visibility in the continuum ($V_c$), the amplitude of the differential visibility ($\Delta V_{\mathrm{diff}}$) and differential phase ($\Delta \phi_{\mathrm{diff}}$) variations in the line.

To estimate the absolute visibility signal we used the Eq. \ref{eq:vc}. The amplitude of the differential visibility variation in the line is given by:
 \begin{equation}
\Delta V_{\mathrm{diff}}= 1-V_{\mathrm{*}}/V_{c} = \frac{1}{V_c}\frac{V_c F_c+ V_{l} F_l}{F_c+F_l} ,
\label{eq:vdiff}
 \end{equation}
where $V_{\mathrm{*}}$ is the source visibility, $V_{l}$ the source visibility averaged over the line, $V_{c}$ the visibility in the nearby continuum and $F_l$  and $F_c$=1 are the line and continuum flux, respectively.

We consider two extreme cases for the differential visibility signal:  $R_{\mathrm{blr}}<<R_{\mathrm{rim}}$ and $R_{\mathrm{blr}}$=2$R_{\mathrm{rim}}$. In the first case, we assume that the BLR is fully unresolved by the interferometer, i.e., $V_{\mathrm{blr}}$=1. Thus, using Eq. \ref{eq:vdiff} for the unresolved BLR (i.e. $V_{l}$=1) we obtain:
\begin{equation}
\Delta V_{\mathrm{diff}} = -\frac{F_l}{1+F_l} \frac{\alpha_c^2}{1-\alpha_c^2} .
\end{equation}
For the large BLR case, the line visibility can be written as $V_{l}$ = 1 - 2 (2$\alpha_c$)$^2$, consequently:
	
\begin{equation}
\Delta V_{\mathrm{diff}} = 3 \frac{F_l}{1+F_l} \frac{\alpha_c^2}{1-\alpha_c^2} .
\end{equation}
Finally, the typical differential phase amplitude for the BLR is given by
 \begin{equation}
 \Delta\phi_{\mathrm{diff}}=\pi \frac{F_l}{1+F_l}\alpha_{l}\mathrm{cos}\,\omega,
 \label{eq:phi1}
 \end{equation}
 where $\alpha_{l}=\frac{2R_{\mathrm{blr}}}{\lambda/B}$.
If the inner rim of the dust torus is inclined and skewed, differential interferometry will also be sensitive to the difference between the global line and the continuum apparent photocenter with maximum amplitude of
 \begin{equation}
 \Delta\phi_{\mathrm{diff}}\simeq \frac{\pi}{2} \frac{F_l}{1+F_l} \alpha_c \mathrm{sin}\,i.
 \label{eq:phi2}
 \end{equation}

 \subsection{Selection of targets} \label{sec:target}
 %Insert Figure============================================
 
 We collected a list of all Sy1 and QSOs observable with the VLTI found in the SIMBAD catalog with search criteria $K_{\mathrm{mag}}<13$, $V_{\mathrm{mag}}<15$ and dec $<15\degree$. For each source we estimate the inner rim radius from its magnitude thanks to an extrapolation of \citet{2006ApJ...639...46S} known measurements. From this rim radius we evaluate the possible values of the continuum visibility, differential visibility and phase. These values are compared to the SNR estimates deduced from the source estimated $K$ magnitude. We use the CMB corrected redshift for each target from NED and $K$ magnitude from 2MASS point source catalog. We collected the list of objects from \citet{2013ApJ...767..149B} that have classical RM BLR size. Then we fitted the radius with their $K$ magnitude and extrapolate for the objects that do not have the RM BLR size.

For each target, we use the strongest emission line in the $K$-band given the actual redshift of the target. To compute the interferometric observables we used the following values for the normalized line flux $F_l$.
 \begin{itemize}
 \item $F_l=0.6$ when $\mathrm{Pa}\, \alpha$ is in the $K$-band ($0.08\le z < 0.25$)
 \item $F_l=0.3$ when $\mathrm{Pa}\, \beta$ is in the $K$-band ($0.4\le z<0.87$)
 %\item $F_l=0.3$ when $\mathrm{Pa}\, \beta$ is in the $H$-band ($0.25\le z<0.4$)
 \item $F_l=0.06$ when $\mathrm{Br}\, \gamma$ is in the $K$-band ($z<0.08$)
 \item $F_l=0.12$ when $\mathrm{Pa}\, \gamma$ is in the $K$-band ($z\ge 0.87$)
 \end{itemize}
 These mean values are deduced from the IR line intensity measurements in \citet{2008ApJS..174..282L}. The dispersion of line strengths is very large. For example the $\mathrm{Br}\, \gamma$ line flux goes from 0.01 to 0.18 with a 0.07 mean. This limited dataset does not allow good statistics but we used it to estimate that about one third of the targets where $\mathrm{Br}\, \gamma$ must be used, will eventually be impossible to observe ($F_l<0.02$)  while our signal estimates are actually pessimistic for half of the targets where $F_l$ is larger than the mean values used here.
 
 %The $K_{\mathrm{mag}}$ of each object is corrected taking the $K_{\mathrm{mag}}$ from 2MASS point source catalog subject to availability. We subtracted the contribution of host galaxy, which is taken 0.2 in $K_{\mathrm{mag}}$ \citep[see, last paragraph of introduction in][] {2004ApJ...600L..35M}. 

 \subsection{Feasibility of observation}
Fig. \ref{Fig:target} summarizes the feasibility of observation of BLRs with current, upcoming, and possible VLTI instruments. We found that AMBER+ would observe at most a few sources whereas the GRAVITY will provide high quality differential measurements on 10 to 15 sources for which it would also give decisive absolute visibility measurement. On the other hand, OASIS would at least double the number of differential phase measurement (with respect to GRAVITY). Moreover, OASIS+ would again double this accessible number of targets then the OASIS number. A new generation FT would boost the GRAVITY list of targets. The ultimate VLTI performance would be obtained with the new generation FT and OASIS+. Remember that an instrument accessing more targets also gives much better results on the brighter targets.

%It is obvious that AMBER+ in the current version cannot observe more than a few BLRs, and only in differential mode. On the other hand, we found that improved version of AMBER like OASIS and OASIS+ could observe tens of objects, for which we can get differential interferometric measurements, and a few objects on which absolute visibility could be determined.
%The upcoming instrument GRAVITY with its new fringe tracking capability up to $K_{\mathrm{mag}}=9.5$ for Sy 1 AGNs and $K_{\mathrm{mag}}=10.5$ for QSOs is likely to have intermediate capabilities between AMBER and its possible upgrade, providing differential measurements on about 10 to 20 targets and absolute visibility on less than 10.   

\begin{table*}	
	\centering
	\caption{Simulated data with fixed $\sigma_0=85\, \mathrm{km/s}$}\label{table:3}
	\begin{tabularx}{\linewidth}{*{6}{p{.15\linewidth}}}
	\multicolumn{6}{c}{True parameters} \\ \toprule
		     Dataset  & instrument  &  $\sigma_{\mathrm{\mathrm{blr}}}$ (mas) & $\mathrm{log}(M_{\mathrm{bh}}/M_{\mathrm{sun}})$ & $i(\degree)$  & $\omega (\degree)$  \\ \midrule
		     A  & AMBER+ & 0.4 &  8 & 30 & 40   \\
		     B  & GRAVITY & 0.4 & 8  & 30  & 40   \\
		     C  & AMBER+ & 0.4 &  8 & 15  &  60   \\ \bottomrule        
     \end{tabularx}
     
	\bigskip

    \begin{tabularx}{\linewidth}{*{5}{p{.19\linewidth}}}
		\multicolumn{5}{c}{Recovered parameters} \\ \toprule
		     Dataset  & $\sigma_{\mathrm{\mathrm{blr}}}$ (mas) & $\mathrm{log}(M_{\mathrm{bh}}/M_{\mathrm{sun}})$ & $i(\degree)$  & $\omega (\degree)$  \\ \midrule
		     A  & $0.378^{+0.015}_{-0.010}$ &  $8.059^{+0.126}_{-0.123}$ & $24.8^{+5.4}_{-3.5}$ & $39.2^{+9.0}_{-7.3}$   \\
		     B  & $0.379^{+0.017}_{-0.015}$ &  $8.020^{+0.085}_{-0.056}$ & $28.5^{+2.0}_{-2.8}$ & $37.2^{+4.7}_{-4.7}$   \\
		     C  & $0.386^{+0.026}_{-0.024}$ &  $7.997^{+0.106}_{-0.076}$ & $15.7^{+4.4}_{-3.2}$ & $58.7^{+6.9}_{-9.4}$  \\ \bottomrule
    \end{tabularx}
\label{TABLE:f_dv}	
\end{table*}

\begin{table*}	
	\centering
	\caption{Simulated data with fixed $\sigma_{\mathrm{\mathrm{blr}}}=0.4$ mas}\label{table:4}
	\begin{tabularx}{\linewidth}{*{6}{p{.14\linewidth}}}
	\multicolumn{6}{c}{True parameters} \\ \toprule
		Dataset  & instrument  &  $\mathrm{log}(M_{\mathrm{bh}}/M_{\mathrm{sun}})$ & $i(\degree)$  & $\omega(\degree)$ & $\Delta v_0(2.35\, \sigma_0) \mathrm{km/s}$ \\ \midrule
		D  & AMBER+ &  8 & 30 & 40 & 500  \\
		E  & AMBER+ &  8 & 30 & 40 & 1500  \\
		F  & AMBER+ &  8 & 10  &  60  & 1500   \\ \bottomrule
	\end{tabularx}
	
	\bigskip
	
	\begin{tabularx}{\linewidth}{*{5}{p{.18\linewidth}}}
	\multicolumn{5}{c}{Recovered parameters} \\ \toprule
		 Dataset  & $\mathrm{log}(M_{\mathrm{bh}}/M_{\mathrm{sun}})$ & $i(\degree)$  & $\omega(\degree)$ & $\Delta v_0(2.35\, \sigma_0) \mathrm{km/s}$ \\ \midrule
		  D  & $8.008^{+0.088}_{-0.059}$ &  $28.3^{+3.0}_{-2.8}$ & $36.6^{+4.9}_{-4.2}$ & $684.8^{+86.8}_{-137.8}$   \\
		  E  & $8.138^{+0.060}_{-0.065}$ &  $20.8^{+5.4}_{-2.4}$ & $40.9^{+3.3}_{-6.6}$ & $1324.2^{+172.4}_{-169.7}$   \\
		  F  & $8.004^{+0.134}_{-0.085}$ &  $18.5^{+9.3}_{-5.5}$ & $58.0^{+8.8}_{-18.4}$ & $1481.2^{+81.4}_{-171.5}$ \\ \bottomrule
	\end{tabularx}	
	\label{TABLE:f_sigma}
\end{table*}

 \section{Parameter uncertainty from Simulated data}
 In section §\ref{sec:Obs_signatures} we described the observables signatures of the main BLR parameters and in section §\ref{sec:present and future} we evaluated signal to noise ratio of OI measurements. Eventually the BLR model parameters will be estimated from a global model fit of RM and OI observables. In this section we give a first estimate of the accuracy of some parameters after a global fit of OI observables with the SNR of a few typical observations.
 % of some of the parameters with typical observing parameters.
  
 \subsection{Simulated datasets}
 To estimate uncertainty of the parameters from optical interferometry data, we created mock datasets using the values mentioned in Table \ref{TABLE:f_dv}a and Table \ref{TABLE:f_sigma}a, considering Gaussian noise on all the spectro interferometric observables. We considered AMBER+ and GRAVITY with absolute visibility accuracy of 3\% and 0.5\% respectively. For AMBER+ we considered $\sigma_{\phi_{D}}$=0.01 radian and $\sigma_{V_D}=\sqrt{2}\times \sigma_{\phi_{D}}$ (using \citet{1986JOSAA...3..634P} relation between the accuracy of differential visibility and phase set by fundamental noise for unresolved sources). For GRAVITY we took $\sigma_{\phi_{D}} \simeq 0.002$ radian. We considered 0.2 \% uncertainty on the line flux measurement.   
 
 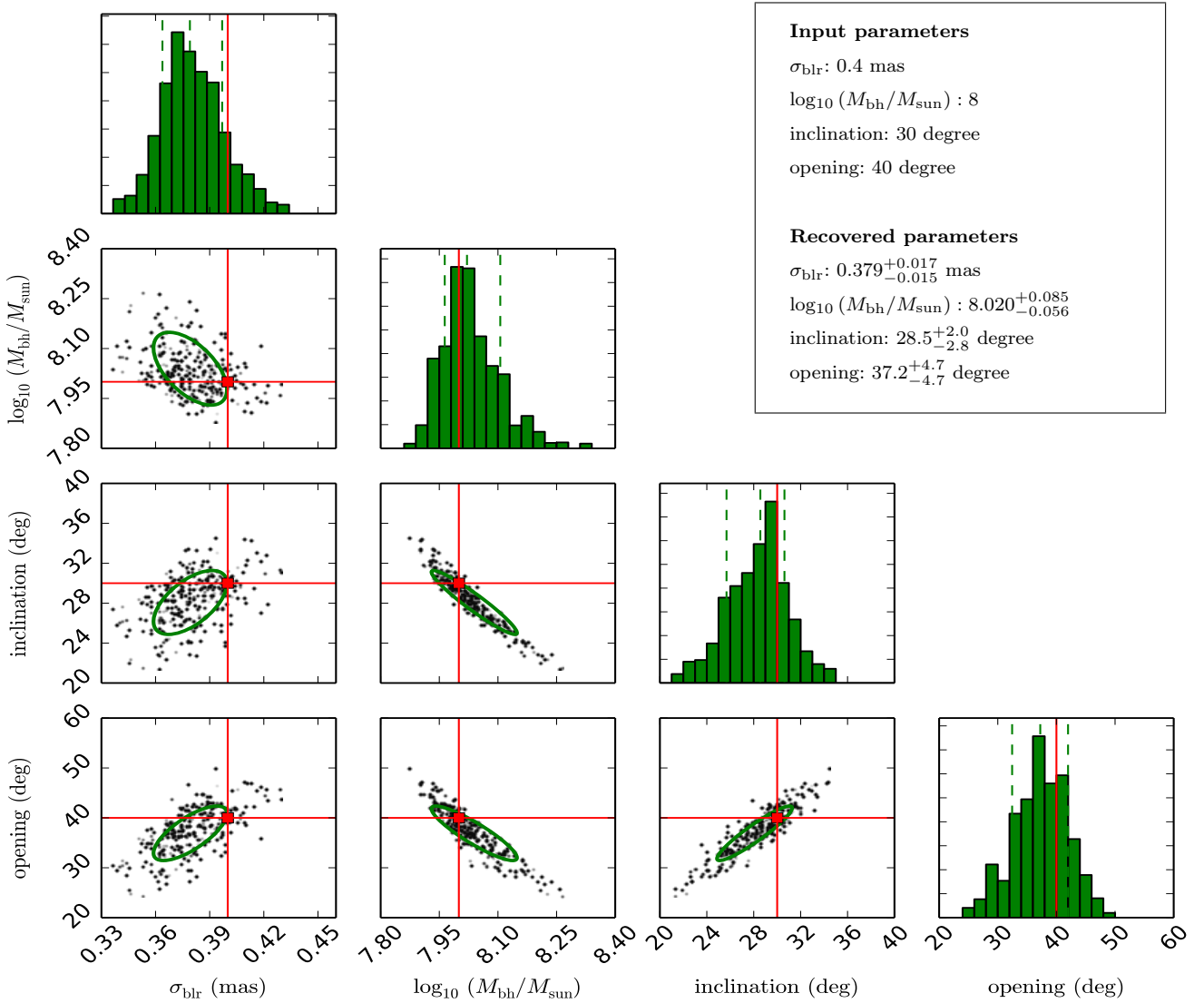
\begin{figure*}
 \setlength{\unitlength}{1cm}
     \begin{picture}(18, 15)
     \put(0,0){\includegraphics[width=18cm,height=15cm]{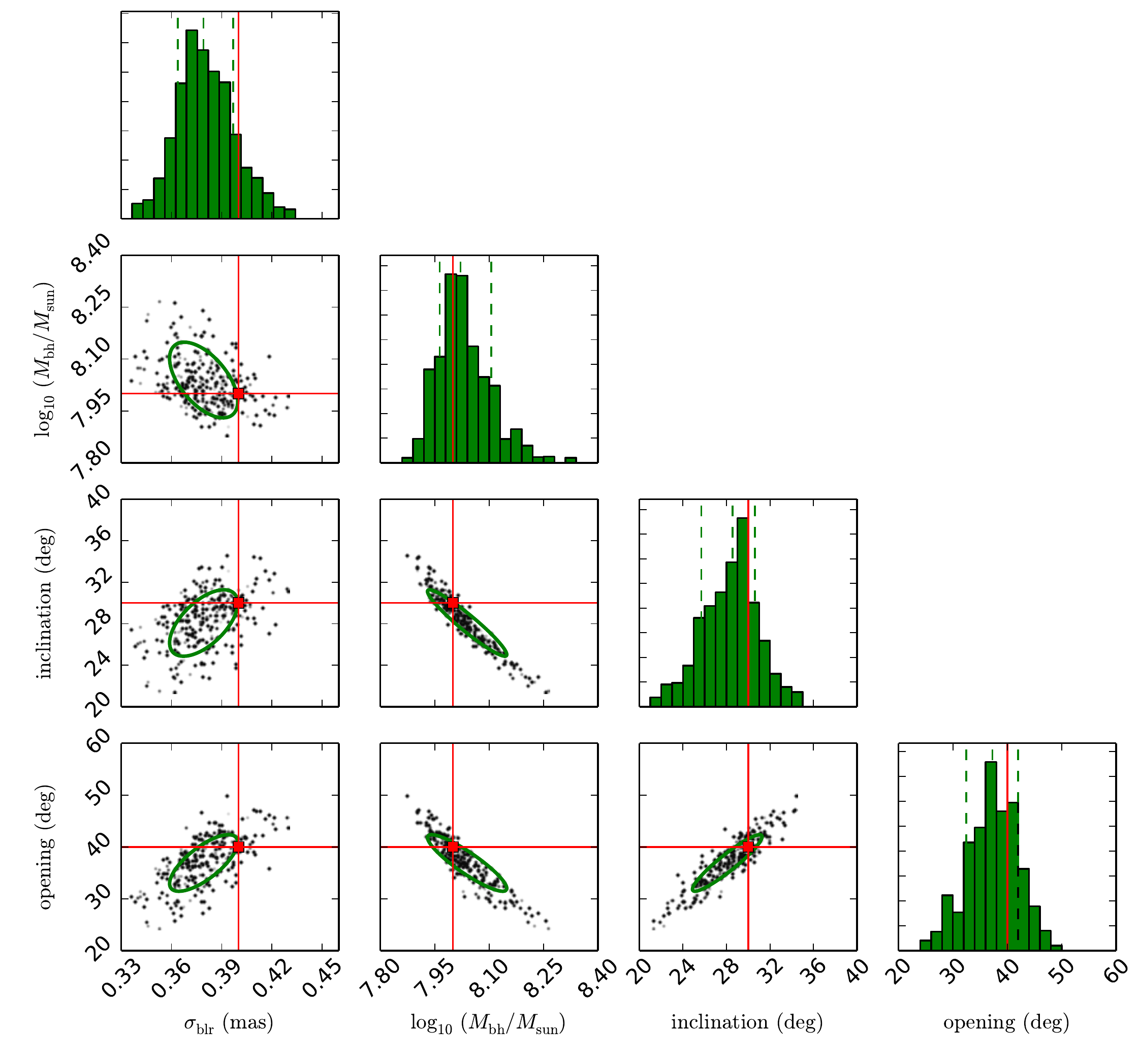}}
     
     \put(11.5,9){\framebox(6,6){}}
     \put(12, 14.5) {{\bf Input parameters}}
     \put(12, 14) {$\sigma_{\mathrm{blr}}$: 0.4 mas}
     \put(12, 13.5) {$\mathrm{log_{10}}\,(M_{\mathrm{bh}}/M_{\mathrm{sun}}):8$}
     \put(12, 13){inclination: 30 degree}
     \put(12, 12.5){opening: 40 degree}
     
     \put(12, 11.5) {{\bf Recovered parameters}}
     \put(12, 11) {$\sigma_{\mathrm{blr}}$: $0.379^{+0.017}_{-0.015}$ mas}
     \put(12, 10.5) {$\mathrm{log_{10}}\,(M_{\mathrm{bh}}/M_{\mathrm{sun}}):8.020^{+0.085}_{-0.056}$}
     \put(12, 10){inclination: $28.5^{+2.0}_{-2.8}$ degree}
     \put(12, 9.5){opening: $37.2^{+4.7}_{-4.7}$ degree}
     
     \end{picture}
     \caption{EMCMC post-burn distributions for dataset B. The red line shows actual input parameters of this dataset. The scatter plots show the projected two-dimensional distributions and green ellipses represents the one $\sigma$ regions of the projected covariance matrix. The histograms show the projected one-dimensional distributions with dotted green lines representing mean and the one $\sigma$ uncertainties. From top-to-bottom and left-to-right, the panels show BLR width $\sigma_{\mathrm{blr}}$, $\mathrm{log_{10}}\,(M_{\mathrm{bh}}/M_{\mathrm{sun}})$, inclination $i$ and opening angle $\omega$.} \label{Fig:emcee_Gravity}
\end{figure*}

 \subsection{Recovering parameters from Simulated datasets}
  In order to recover the parameters of the datasets and their associated uncertainties, we optimized the likelihood function that we considered to be Gaussian and defined by
  \begin{equation}
  p(\mathrm{data|model}) = \prod_{i=1}^N \frac{1}{\sqrt{(2\pi \sigma_i^2)}} \mathrm{exp}\left[-\frac{(\mathrm{data}_i - \mathrm{model}_i)^2}{2\sigma_i^2}\right],
  %p(data|model)\propto exp\left[-\frac{1}{2}\chi^2(data, model)\right].
  \end{equation} 
  where $\sigma_i$ is the uncertainty on $\mathrm{data}_i$. Maximizing the likelihood is identical to minimizing the $\chi^2$. We considered all OI observables i.e. spectrum, differential visibility, differential phase and absolute visibility are the part of our datasets and minimized the global $\chi^2$. 
  According to the Bayes' theorem, the posterior probability distribution $p(\mathrm{model|data})$ is linked with the prior function $p(\mathrm{model})$ which includes any previous knowledge about the parameters:
  \begin{equation}
  p(\mathrm{model|data}) \propto p(\mathrm{model}) \times p(\mathrm{data|model}). 
  \end{equation}
   We assigned uniform prior to all the parameters except black hole mass for which we used log uniform prior. 
   
   To sample the parameters efficiently we used EMCEE package, developed by \citet{2013PASP..125..306F}, which is Python implementation of Affine Invariant Markov Chain Monte Carlo (MCMC) ensemble sampler by \citet{2010AMCS..5..1G}.
   %For the efficient sampling of the parameters space we turn to Markov Chain Monte Carlo (MCMC) analysis using EMCEE package developed by \citet{2013PASP..125..306F}, which is python implementation of Affine Invariant MCMC ensemble sampler by \citet{2010AMCS..5..1G}. 
  EMCEE explores the full posterior distribution using set of random walkers in each step and the result of the walkers is used for the next step in order to optimize the maximum likelihood. We run EMCEE with 200 walkers and 200 steps. After few steps the samples converge and we consider 150 steps as “burn-in” phase. Final 50 steps are considered as post burn-in phase and used to estimate the parameters and the uncertainties.

 An example of the post burn distribution of samples is shown in Fig. \ref{Fig:emcee_Gravity}, which is obtained for dataset B. The scatter plots show the 2D distribution of samples with one $\sigma$ ellipse representing the covariance matrix whereas the histograms show 1D cut of the samples. The ellipses indicate moderated degeneracy of the parameters $M_{\mathrm{bh}}$-$i$ and $M_{\mathrm{bh}}$-$\omega$ as well as $i$-$\omega$, which globally underline the critical sensitivity to the inclination $i$. For all datasets, the recovered parameters and their one $\sigma$ uncertainties are given in Table \ref{TABLE:f_dv} and Table \ref{TABLE:f_sigma}. Most of the parameters are recovered within one $\sigma$ uncertainty. The maximum uncertainty in $\sigma_{\mathrm{blr}}$ is obtained in the case of dataset C, which is $0.386^{+0.026}_{-0.024}$ mas. Inclination has maximum uncertainty $24.8^{+5.4}_{-3.5}$ degree, which is found in the case of dataset A. Opening angle is constrained well in all datasets and one $\sigma$ uncertainty is less than 10 degree. 
 
 As discussed in section §\ref{sec:BLRsize}, OI can provide a semi-independent estimate of the BLR size that can also be deduced from the RM typical time lag. Thus we can concentrate on the degeneracy between $i$, $\omega$ and $\sigma_0$ that impacts on the mass measurement. The results of a model fit with a fixed $\sigma_{\mathrm{blr}}$ are given in Table \ref{TABLE:f_sigma}. 
 
 From a fit of the OI data only we obtain a good constraint on all the parameters with AMBER+ quality level, with a mass accuracy between $\sim 0.08$ and $\sim 0.13$ dex. The largest value is obtained for a quite low $10\degree$ inclination. This uncertainty is quite similar to that achieved by \citet{2011ApJ...730..139P} with simulated RM data. However, when \citet{2012ApJ...754...49P} fit the real RM data of Mrk 50, they found a much larger uncertainty that they attribute to the modeling error. Remember that the statistical uncertainty of 0.15 dex obtained in traditional RM result \citep{2009ApJ...697..160B,2010ApJ...721..715D} neglects the scatter of 0.44 dex in the RM scale factor $f$ \citep{2010ApJ...721...26G,2010ApJ...716..269W}. In that context, our 0.08-0.13 dex results dealing specifically with the major causes of $f$ dispersion are very encouraging, even if it is a minimum value because this first global fit of OI data uses a very simplified model and a quite limited number of parameters. The more accurate measurements of GRAVITY than AMBER+ will bring very substantial progress, as already indicated by dataset B. A global fit of OI and RM data will further improve accuracy of the parameters.

 \section{Discussion, Conclusion and Perspectives}\label{sec:diss}
  This paper is the first in a series on the application of OI to observe QSO BLRs that have or can also be observed with RM. We have presented a three-dimensional model that self-consistently infers the BLR structure from differential interferometry and predicts the reverberation mapping observables. We have restricted ourselves to a limited set of parameters, as our first goal was to understand the typical OI signatures of the BLR features and to evaluate the potential of the QSO BLR observations with the VLTI. However, our model can be very easily updated by changing the properties of the list of clouds making up the BLR, for example with a different radial distribution or by forcing the clouds to be located on a specific surface such as in bowl shaped BLR models. 
  
  We show that OI, with a spectral resolution of the order of 1000, will remove the degeneracies between the inclination $i$, the opening angle $\omega$ and the local velocity field contribution $\sigma_0$ that are the main cause for the dispersion of mass estimates from RM using only the equivalent time lag and width of the emission lines. Monte Carlo Markov Chain (MCMC) model fits with the EMCEE package applied to the simulated differential visibility and phase datasets confirmed that OI alone can measure the BH mass, $i$, $\omega$ and $\sigma_0$, if we have a good estimate of the BLR angular size. The resulting mass estimate accuracy will be of the order or better than 0.08 dex with AMBER+, (except at inclinations lower than $10\degree$), and will be further improved with GRAVITY. This is much better than the standard mass dispersion of 0.30 to 0.44 dex that includes the effect of $f$ dispersion \citep{2009ApJ...697..160B,2010ApJ...716..269W}. It is similar to the best advanced model fits of RM data by \citet{2011ApJ...730..139P,2014arXiv1407.2941P}. Combining OI and RM  will validate these model fits and increase the number of usable equations and therefore the number of parameters that  can be fitted and well separated.
    
  We have underlined the importance of high accuracy absolute visibility measurements. A key condition is the possibility to use a fringe tracker that stabilizes the OI transfer function and reduces its calibration errors. However, there are many targets where absolute visibility measurements will not be accurate enough while we will still have accurate differential visibilities and phases. Then it will be necessary to obtain the size information from RM measurements, but this assumes that we can compute the scaling factor between OI and RM sizes discussed in section \ref{sec:BLRsize} as well as the scaling factor between OI and RM observations made in different emission lines. We will come back to this key point at the end of this discussion.
  
  When the global angular size of the BLR has been estimated, either from direct OI observables or from properly scaled RM observables, differential visibilities and phases, or even differential phases alone, are sufficient for accurate mass estimates, if the interferometric observations feature a sufficient SNR. In addition to $i$, $\omega$ and $\sigma_0$, these observables constrain the other BLR characteristics such as the nature of the global velocity field (rotation and inflow-outflow velocity laws) or the cloud optical thickness.
  
  To evaluate the potential of interferometric observations of QSO BLRs with the current and near future VLTI instruments, we have computed the expected accuracy for absolute visibility, differential visibility and phase with current (AMBER+), near future (GRAVITY) and possible (OASIS, OASIS+ and OASIS+FT) VLTI instruments. This SNR analysis has been checked on our real 3C273 data from AMBER+ and the values for the other instruments are deduced from elementary cross-multiplications based on the known changes in detector noise, number of pixels, transmission and exposure time of the new instruments. We have considered the possible SNR for all QSOs and Seyfert 1 AGNs observable at Paranal brighter than $K=15$ that is the potential limit for VLTI observations with OASIS+. 
  
  Even if all these BLRs will remain quite unresolved with the VLTI in the $K$-band, we see that measurements are nevertheless possible on many targets. GRAVITY, limited by its internal fringe tracker at $K=10.5$, will give absolute and differential measurements on half a dozen sources. For about fifteen sources we will have only absolute visibility (but no differential visibility) and differential phase that still allows to fit all parameters with some loss in accuracy. OASIS would allow 50\% increase in the number of targets with absolute visibility and differential phase and would double the number of sources with differential phase measurements. OASIS+ would yield all observables on more than 40 targets. A fringe tracker reaching $K\simeq13$ would be a major breakthrough by extending the number of GRAVITY targets to about 30 and of OASIS+ targets to about 60 and by improving the accuracy on all measurements. This seems well within the reach of the currently proposed designs \citep{2012SPIE.8445E..1LM, 2014SPIE.9146E..2P}. The full sample of VLTI targets would allow trying a general unification of BLR model by studying for example the cross-relations of the key parameters such as the projection factor $f$, the BLR thickness $\omega$, the local velocity field parameter $\sigma_0$ or the rotation-to-inflow velocity ratio as a function of the luminosity. The VLTI, with its full potential, could allow exploring four to five decades of luminosity range. Full imaging of BLRs requires improving the angular resolution by a factor at least ten that requires a major breakthrough on sensitivity of OI in the visible, on CHARA for example, or the construction of a new interferometer with larger baselines. These are long term goals, while differential interferometry of QSOs with the VLTI has already started with AMBER , will substantially expand very soon with GRAVITY and can reach its full potential with a new generation fringe tracker and a specialized small instrument like OASIS+ in quite less than 5 years.
   
  This paper concentrates on the typical signatures of the BLR parameters in optical interferometric measurements, and on the improvement of the black hole mass estimates. The next papers in our series will treat the issue of distance measurements from combined reverberation mapping and optical interferometric measurements and, later the improvement and hopefully unification of BLR and dust torus models. The combination of optical interferometry and reverberation mapping measurements raises the problem of the scaling factors between the BLRs seen in different hydrogen lines. Indeed, reverberation mapping is currently available in the visible while optical interferometry with large apertures is developing first in the near-IR. 
  Eventually, there will be reverberation mapping measurements in the $\mathrm{Pa}\, \alpha$ and $\mathrm{Pa}\, \beta$ as discussed for example by \citet{2013MNRAS.432..113L} who shows that near-IR offers several advantages over the visible, but it would take some time to accumulate a relevant quantity of data. On the other hand, OI in the visible might reach the sensitivity needed for some QSO BLR observations around 2020.
  
  For the VLTI applications of the next years, we still have to estimate the scaling factor between different lines. It should be relatively easy to obtain a good approximation from the application of a code like Cloudy \citep{1998PASP..110..761F,2013RMxAA..49..137F} to the known global geometry of the cloud distribution given by DI alone. Then, it will be possible to combine OI and RM, and to have a better modeling of the scale factor converging iteratively toward the right combination of RM in the visible and DI in the IR. This will be a crucial step, in particular for direct distance estimates from the combination of angular and linear measurements. This study will be the main focus of the second paper in that series concentrating on the measurement of QSO distances and on the calibration of the size-luminosity law, in order to allow QSOs to be used as standard cosmological candles. Another aspect is the combination of BLR observation with the torus observations and in some cases images of the innermost parts of the dust torus with the second generation VLTI instrument MATISSE \citep{2012SPIE.8445E..0RL}. This will require some substantial modeling to study the dust structure at various wavelengths and use this information to complement the available differential observation of BLR in lines. Thus it will be critical that MATISSE observes all sources on which we can obtain BLR data, with MATISSE itself in the $L$-band but also with all $K$-band instruments.

\section{ACKNOWLEDGEMENT}

We thank Walter Jaffe (Leiden University) for a decisive discussion about SNR in coherent and incoherent integration modes, Stephane Lagarde (OCA, Nice) for its initial help with the modeling of interferometric observables, Florentin Millour (OCA, Nice) for its contribution to AMBER+, Frantz Martinache (OCA, Nice) for reading and improving the manuscript, Makoto Kishimoto (Kyoto University), Alessandro Marconi (University of Florence) and Gerd Weigelt (MPIfR, Bonn) for enlightening discussions about OI and AGNs and Daniel Moser Faes (Universidade de S\~{a}o Paulo) for useful discussion about EMCEE. 

This research used the SIMBAD database, which is operated at CDS, Strasbourg, France, the NASA/IPAC Extragalactic Database (NED), which is operated by the Jet Propulsion Laboratory, California Institute of Technology, under contract with NASA, and data products from the Two Micron All Sky Survey, which is a joint project of the University of Massachusetts and the Infrared Processing and Analysis Center/California Institute of Technology, funded by the NASA and the National Science Foundation. 

SR is supported by the Erasmus Mundus Joint Doctorate Program by Grant Number 2011-1640 from the EACEA of the European Commission. 

SR also thanks Jaba and Sunil Rakshit and Neha Sharma (ARIES, India) for their support and encouragement during this work. 

%******************************************************************************************
\bibliographystyle{mn2e}
\bibliography{ref}
%******************************************************************************************

%%%%%%%%%%%%%%%%%%%%%%%%%%%%%%%%%%%%%%%%%%%%%%%%%%%%%%%%%%%%%%%%%%%%%%%%%%%%%%%%%%%%%%%%%%%
\label{lastpage}
%%%%%%%%%%%%%%%%%%%%%%%%%%%%%%%%%%%%%%%%%%%%%%%%%%%%%%%%%%%%%%%%%%%%%%%%%%%%%%%%%%%%%%%%%%%
\end{document}